\documentclass{aa}  

\usepackage{graphicx}
\usepackage{txfonts}
\usepackage{hyperref}
\usepackage{orcidlink}
\usepackage{xcolor}
\definecolor{mylightblue}{rgb}{0.19,0.55,0.91}
\hypersetup{colorlinks,breaklinks,
            urlcolor=mylightblue,citecolor=violet,
            anchorcolor=lightgray,linkcolor=mylightblue}
\graphicspath{{figures/}{../figures/}}

\begin{document} 

\def\teff{$T_{\rm eff}$}
\def\gr{$\log {\rm g}$}

   \title{The earliest phases of CNO enrichment in galaxies\thanks{Full Tables~1 and 2 are only available at the CDS via anonymous ftp
to \url{cdsarc.u-strasbg.fr (130.79.128.5)} or via \url{http://cdsarc.u-strasbg.fr/viz-bin/cat/J/A+A/xxx/Ayyy}}}
   
   \subtitle{}

   \author{ 
          M. Rossi \inst{1,2}\orcidlink{0000-0001-6887-2663}
          \and
          D. Romano\inst{2}\orcidlink{0000-0002-0845-6171}
          \and
          A. Mucciarelli\inst{1,2}\orcidlink{0000-0001-9158-8580}
          \and 
          E. Ceccarelli\inst{1,2}\orcidlink{0009-0007-3793-9766}
          \and 
          D. Massari\inst{2}\orcidlink{0000-0001-8892-4301}
          \and
          G. Zamorani\inst{2}\orcidlink{0000-0002-2318-301X}
          }

   \institute{Dipartimento di Fisica e Astronomia, Alma Mater Studiorum, Università di Bologna, Via Gobetti 93/2, 40129 Bologna, Italy\\
              \email{martina.rossi64@unibo.it}
         \and
             INAF, Osservatorio di Astrofisica e Scienza dello Spazio, Via Gobetti 93/3, 40129 Bologna, Italy
             }

   \date{Received June 20, 2024; accepted Oct 3, 2024}

   \titlerunning{The earliest phases of CNO enrichment}
   \authorrunning{Rossi et al.}

 
  \abstract
   {The recent detection of  {super-solar carbon-to-oxygen and nitrogen-to-oxygen abundance ratios in a bunch of} metal-poor galaxies at high redshift by the {\it James Webb} Space Telescope has sparked renewed interest in exploring the chemical evolution of carbon, nitrogen, and oxygen (the CNO elements) at early times, prompting fresh inquiries into their origins.} 
   {The main goal of this paper is to shed light onto the early evolution of the main CNO isotopes in our Galaxy and in young distant systems, such as GN-z11 at $z = 10.6$  {and GS-z12 at $z = 12.5$}.}
   {To this aim, we incorporate a stochastic star-formation component into a chemical evolution model calibrated with high-quality Milky Way (MW) data, focusing on the contribution of Population~III (Pop~III) stars to the early chemical enrichment.}
   {By comparing the model predictions with CNO abundance measurements from high-resolution spectroscopy of an homogeneous sample of Galactic halo stars, we first demonstrate that the scatter observed in the metallicity range $-4.5 \le$~[Fe/H]~$\le -1.5$ can be explained by pre-enrichment from Pop~III stars that explode as supernovae (SNe) with different initial masses and energies. Then, by exploiting the chemical evolution model, we provide testable predictions for log(C/N), log(N/O), and log(C/O) vs. log(O/H)+12 in MW-like galaxies observed at different cosmic epochs/redshifts. Finally, by calibrating the chemical evolution model to replicate the observed properties of GN-z11  {and GS-z12}, we provide an alternative interpretation of  {their high N/O and C/O abundance ratio, respectively,} demonstrating that  {an anomalously high N or C} content can be reproduced through enrichment from faint Pop~III SNe.}
   {Stochastic chemical enrichment from primordial stars explains both the observed scatter in CNO abundances in MW halo stars and the exceptionally high  {C/O and} N/O ratios in some distant galaxies. These findings emphasize the critical role of Pop~III stars in shaping early chemical evolution.}


   \maketitle
%

\section{Introduction}
\label{intro}

Comparing chemical abundance measurements in gas and stars with the predictions of galactic chemical evolution (GCE) models allows to place tight constraints on the modes and time scales of the assembly of galactic (sub)structures, as well as on stellar evolution and nucleosynthesis theories. In particular, analyzing the median or mean trends of chemical abundance ratios for elements originating from different nucleosynthetic channels serves as a diagnostic tool for understanding various aspects such as star formation efficiency, gas accretion, outflows, and formation history of galaxies  {\citep[e.g.,][]{Griffith19, Molla19, Spitoni2019, Spitoni23, Weinberg19, Hasselquist21, Matteucci2021}}. At the same time, the dispersion of data points around these mean trends provides further insights. Indeed, the scatter around the mean trends can stem from variations in the relative contributions of different stellar polluters, from the stochastic star formation process itself, or from both.
At low metallicities, $\rm [Fe/H] < -2$, a significant increase of the star-to-star scatter is observed for different elements \citep[e.g.,][]{Griffith23}.  This evidence has been interpreted as a distinctive indicator of the presence of  stochastic enrichment processes \cite[e.g.,][]{Argast2004,Welsh21} along with the sign of the chemical enrichment produced by the first stars \citep[e.g.,][]{SS07,ioanna2023,Vanni23}.

Strikingly, the formation and evolution of the first stars that are born from primordial matter free of metals, the so-called Population~III (Pop~III) stars, remain shrouded in mystery. In fact, despite many observational campaigns dedicated to identifying their offspring, we are still far from having a clear picture of their initial mass distribution and evolutionary pathways. It has been suggested that Pop~III stars are predominantly massive (and, hence, short-lived), with initial masses tens to hundreds, or even thousands, of times that of the Sun, depending on the interplay between several processes --cloud collapse, dynamic gas accretion, protostellar disc fragmentation, stellar feedback-- with the effects of magnetic fields yet to be fully explored \citep{Bromm1999,Bromm2002,Abel2002,Hosokawa2011,Hosokawa2016,Sugimura2020,Saad2022,Sharda2024}. As shown in simulations, long-lived, low mass Pop~III stars can also form \citep[e.g.,][]{Clark2011,Latif2022}, albeit at a sensibly reduced pace compared to typical present-day star formation conditions.

On the above theoretical premises, it is clear that there is little hope to observe directly primordial (zero-metal) stars in the local universe: the record holder as the star with the lowest proportion of metals is the Galactic halo star SDSS\,J102915+172927 \citep[$Z \le 4.5 \times 10^{-5} \, Z_\odot$,][]{Caffau2011}. Nevertheless, the chemical footprint of the first stars can be found in the atmospheres of the oldest low-mass stars that have survived to the present day \citep{Cayrel2004,Beers2005,Frebel2005,Debennassuti2014,Salvadori2019, ioanna2023, Vanni23}.
Indeed, among ancient very metal-poor stars ($\rm [Fe/H]<-2$), the chemical signatures of Pop~III supernovae (SNe) with different explosion energies have been identified \citep{Iwamoto05, ishigaki14, Ezzeddine2019, Placco21, Skuladottir2021}. 
In the initial mass range 10--100~M$_{\odot}$, Pop~III SNe can release very different explosion energies, ranging from $\rm E_{SN} \approx 10^{50}$~erg to $\rm E_{SN} \approx 10^{52}$~erg (e.g. \citealt{kobayashi06,Heger+woosley10}), while in between 100 and 140~M$_{\odot}$ the stars collapse directly to black holes.
Conversely, Pop~III stars with initial masses within the 140--260 M$_\odot$ range are expected to die as pair-instability supernovae \citep[PISNe,][]{Heger2002}, i.e., thermonuclear explosions following the gravitational collapse triggered by dynamical instability of the CO core. The instability is in turn driven by the production and subsequent annihilation of electron-positron pairs in the core, which occurs when low density ($\le 10^6$~g~cm$^{-3}$) and high temperature ($\ge 10^9$~K) conditions are met \citep{Rakavy1967}. The pair-instability collapse leaves no remnant and restores to the interstellar medium (ISM) all the mass of the star, half of which in the form of newly-produced metals. Stars more massive than 260~M$_\odot$, instead, directly collapse to black holes and do not contribute to the chemical enrichment of the surrounding medium \citep{Heger2002}. While including the effects of rotation and magnetic torques may change the mass range in which PISNe are expected to occur \citep{Yoon2012,Uchida2019}, the crux is that primordial PISNe are expected to imprint a characteristic chemical pattern to the stellar generations that form from their ashes. It has been convincingly demonstrated that long-lived stars born from the gas polluted by the first very massive stars exploding as PISNe should have underabundances of some key elements (N, Cu, Zn) with respect to Fe and [Ca/H]~$\simeq -2.5$, [Fe/H]~$\simeq -2$ \citep{Karlsson2008,Salvadori2019}. Due to this metallicity overshoot, PISN descendants should not be searched among the most metal-poor stars. Below [Fe/H] ~$\simeq -5$, the almost totality of the stars are expected to form from gas uniquely polluted by faint SNe -- the end products of Pop~III stars with masses lower than 100~M$_\odot$. Such faint SNe are characterised by large [C/Fe] ratios in the ejecta, a chemical signature that is washed out at higher metallicities because of the increasing contribution of Pop~II `normal' core-collapse SNe (CCSNe) as time goes by \citep[e.g.,][]{Salvadori2015,Hartwig2018}.

Previous GCE models \citep{Ballero2006} concluded that Pop~III stars have a negligible impact on the early galactic enrichment. Such a conclusion follows from assuming homogeneous mixing of the stellar ejecta over the galactic volume, along with a fully-sampled stellar initial mass function (IMF) for the first stars. However, in order to mimic the formation of the first stellar generations, one should consider localized star formation inside self-enriching gas clumps that later merge together, joined to random sampling of the IMF \citep[e.g.,][]{Argast2004,Rossi+21}.

In this work, we reassess the role of Pop~III stars as early Milky Way (MW) pollutants. In particular, we focus on the evolution of the CNO elements, in view of their importance as gas coolants at low metallicities \citep{Bromm2003,Sharda2023}, alternative H$_2$-gas and star-formation rate tracers \citep[][and references therein]{Narayanan2014,Combes2018,Hunter2024}, and constituents of complex organic molecules that possibly precede the formation of prebiotic species \citep{Herbst2009,Coletta2020,Fontani2022,Rocha2024}, not to mention their importance as diagnostics of mixing and nucleosynthetic processes in stars \citep[][see also \citealt{Romano2022}, for a review]{Charbonnel1998,Gratton2000,Lagarde2024}. Furthermore, we extend our GCE model to high-redshift systems, in which accurate abundances for C, N, and O are being obtained thanks to the Near Infrared Spectrograph  {\citep[NIRSpec,][]{Jakobsen2022}} on board the {\it James Webb} Space Telescope \citep[JWST, see, e.g.,][]{MarquesChaves2024}.

An important, still often overlooked aspect is that of the choice of the proper data set to be used for the comparison with the model outputs. For instance, in stellar spectral analyses the adoption of different stellar model atmospheres, temperature scales, atomic data, and solar reference abundances introduces systematic offsets that are not all easily identifiable and disposable. Uncritically displaying data from different studies all together most likely leads to an artificial widening of the real underlying dispersion, thus hampering the possibility of a truly meaningful comparison with the model predictions. To avoid that, in this work we derive homogeneous CNO abundances for a sample of bona-fide `unmixed' halo stars \citep[][see \citealt{Spite2005} for a definition of the `unmixed' term]{Mucciarelli22}, to which we add further data only after a careful evaluation of all systematics.
Although it may be superfluous to do so, we reiterate here the importance of comparing GCE model predictions for C and N to abundance measurements of stars that have not yet evolved beyond the lower red giant branch (RGB), thus preserving, in principle, the initial CN abundances in their envelopes. Furthermore, binary stars should be excluded to avoid pollution due to mass transfer from an asymptotic giant branch (AGB) companion star \citep{bonifacio15}.

The layout of the paper is as follows. In Section~\ref{sample}, we present the data set used for the comparison with the theoretical predictions. In Section~\ref{models}, we summarize the main characteristics of the adopted GCE model, while in Section~\ref{model implementations} we describe the implementation of the stochastic enrichment from the first stellar generations. The model results for the MW halo are compared to the observations in Section~\ref{resMW}. Section~\ref{reshighz} deals with the extension of the model to the high-redshift universe. Finally, in Sections~\ref{discussion} and \ref{conc}, respectively, we discuss the results and draw our conclusions.

\section{Data sets}
\label{sample}

This section introduces the data used to constrain our models for the early Galaxy (Sect.~\ref{samplehalo}) and young systems detected at high redshift (Sect.~\ref{samplehigh}).

\subsection{The Galactic halo}

\label{samplehalo}
\begin{table*}
    \caption{ID, stellar parameters and chemical abundances of the unmixed halo stars.}
    \centering
    \begin{tabular}{ c c c c c c c c c }
         \hline \hline 
         \noalign{\vskip 0.2cm}

     ID & ID Gaia DR3 & $T_{\rm eff}$ & log$g$ & [Fe/H] &[C/Fe]&[N/Fe] &[O/Fe]& [Mg/Fe]  \\ 
        &             & (K)           & (dex)  & (dex)  &(dex) & (dex) &(dex) & (dex) \\
  \hline
    \noalign{\vskip 0.1cm}

    \multicolumn{9}{c}{\citeauthor{Mucciarelli22}'s (\citeyear{Mucciarelli22}) sub-sample}\\
    \hline
    \noalign{\vskip 0.1cm}

 BD-012582        &   3602201479317188864   &   5224  &   2.82	 &   --2.28$\pm$0.07 &   0.80$\pm$0.10  &  --0.48$\pm$0.12   &  0.65$\pm$0.10  &   0.38$\pm$0.07 \\
CD-241782	 &   5085629204207431296   &   5285  &   2.80	 &   --2.66$\pm$0.07 &   0.24$\pm$0.10  &  --0.67$\pm$0.12   &       --       &   0.40$\pm$0.07\\
CD-30298    	 &   5031398163988267520   &   5274  &   2.71	 &   --3.29$\pm$0.08 &   0.41$\pm$0.08  &    0.14$\pm$0.10   &  $<$1.30        &   0.33$\pm$0.10\\
CS22183-031	 &   2482988183718639872   &   5271  &   2.76	 &   --3.03$\pm$0.08 &   0.45$\pm$0.10  &    0.29$\pm$0.11   &  $<$1.20        &   0.69$\pm$0.06\\
CS22186-023	 &   4868824473488625024   &   5178  &   2.30	 &   --2.64$\pm$0.07 &   0.33$\pm$0.10  &    0.48$\pm$0.13   &  $<$0.90        &   0.29$\pm$0.07\\
\ldots & \ldots & \ldots & \ldots & \ldots & \ldots & \ldots & \ldots & \ldots \\
  \hline
      \noalign{\vskip 0.1cm}

    \multicolumn{9}{c}{\citeauthor{Yong13}'s (\citeyear{Yong13}) sub-sample}\\
    \hline
        \noalign{\vskip 0.1cm}

52972--1213--507 &  811812182598263552 & 6463 & 4.34 & --2.98 & 2.73 &  --   &       --	&       --	\\ 
        53327--2044--515 &   290930261314166528 & 5703 & 4.68 & --4.00 & 1.04 &  --    &       -- &	       --	\\ 
        53436--1996--093 &   767103875148269696 & 6449 & 4.38 & --3.53 &   $< $1.57 &  --   &       --
        &-- 	\\ 
        54142--2667--094 &   598558225898219776 & 6456 & 3.87 & --2.96 &   $< $1.30 &  --    &      -- &       --	\\ 
        BS 16545--089 &   760839819965612928 & 6486 & 3.82 & --3.44 &   $< $1.68 &  --   &       --& -- 	\\ 
\ldots & \ldots & \ldots & \ldots & \ldots & \ldots & \ldots & \ldots & \ldots \\
  \hline
    \end{tabular}
    \tablefoot{The table is available in its entirety at the CDS. A portion is shown here for guidance regarding its form and contents.
    }
    \label{tab spectro}
\end{table*}

Abundances of CNO elements are homogeneously derived for the sample of lower RGB halo stars presented in \cite{Mucciarelli22}. After a careful evaluation of all possible systematics, we complement this sample with the one of \cite{Yong13}. Abundance estimates for dwarf stars from three-dimensional (3D) radiative transfer calculations including corrections for non local thermodynamic equilibrium (non-LTE) conditions \citep{Amarsi19} are added for comparison. We further derive the orbital parameters of the program stars to distinguish those born in situ from the accreted ones (Sect.~\ref{orbital-properties}) and to identify the different progenitors of the latter.

\subsubsection{Stellar Abundances}
\label{sampleale22}

The main sample used in this work is that presented by \cite{Mucciarelli22}, including 58 RGB stars located before the RGB Bump. Therefore, these stars are not affected by the extra-mixing episode associated to this evolutionary feature \citep[e.g.,][]{Cassisi2002,Karakas2014}. For all the targets, 
Fe, C and Li abundances have been already published \citep{Mucciarelli22}.  {In particular, C abundances have been derived from either the G band or the CH feature at 3143~\AA.} For 36 of these targets (covering the metallicity range $\rm -4.5 \le [Fe/H] < -1.0$ dex), the available spectra enabled us to derive N, O and Mg abundances. The spectra used for the analysis were obtained with the Ultraviolet Visual Echelle Spectrograph \citep[UVES,][]{dekker00} at the Very Large Telescope of ESO. Details of the data set are in \cite{Mucciarelli22}.

The adopted atmospheric parameters are taken from \cite{Mucciarelli22}. 
Effective temperatures were derived from the photometry of the early third data release (EDR3) of the ESA Gaia mission \citep{prusti16,brown20}, adopting the ${\rm (BP-RP)_0}$--\teff\  transformation by \citet{mbm21} 
and the colour excess by \citet{ivanova21} and \citet{schlafly11}. Surface gravities were determined using the photometric temperatures, the Gaia parallaxes and the bolometric corrections computed as described in \cite{Mucciarelli22}. Microturbulent velocities have been estimated spectroscopically by minimising any trend between the abundances from Fe~I lines and their reduced equivalent widths.


\begin{figure}
\centering
\includegraphics[width=0.9\columnwidth]{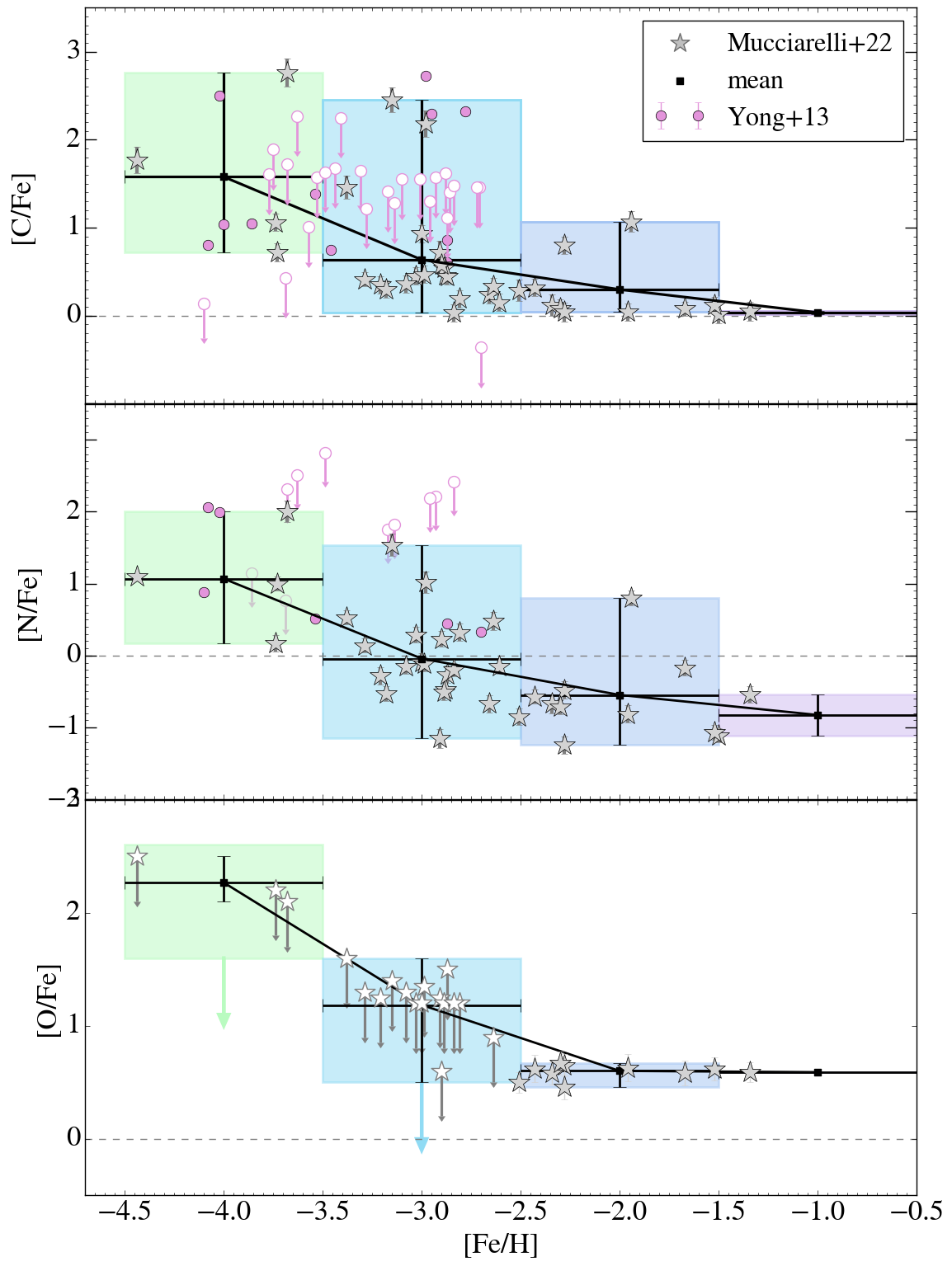}
    \caption{[X/Fe] abundance ratios (where X~= C, N, and O) of the low RGB stars of \citet[][measurements: grey star symbols, upper limits: empty stars]{Mucciarelli22}. Also shown are the mean values (dots connect by solid black lines) and the scatter (coloured boxes) in 1-dex metallicity bins. Adding data from \citet[][purple circles]{Yong13} for C and N increases the scatter, with the caveat that many points are actually upper limits (empty circles).}
    \label{fig1:observed scatter}
\end{figure}


\begin{figure*}
\centering
\includegraphics[width=\textwidth]{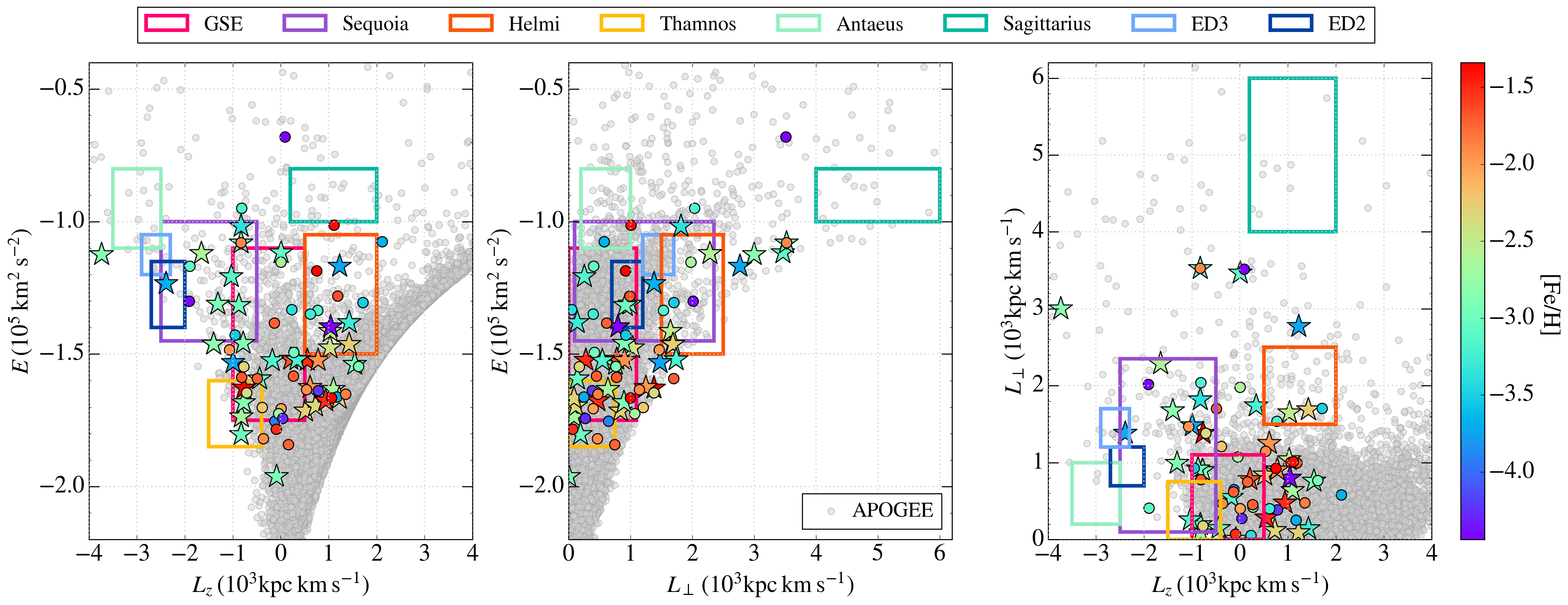}
    \caption{Distribution of the sample stars colour-coded according to their $\rm [Fe/H]$ abundances (circles: stars from \citealt{Yong13}, star symbols: stars from \citealt{Mucciarelli22}) in the IoM spaces. The APOGEE sample is plotted with filled gray points for reference. The boxes represent the positions occupied by previously identified substructures in the Galactic halo \citep{dodd23}.}
    \label{fig:orbital para}

\end{figure*}


\begin{figure*}
\centering

\includegraphics[width=0.75\textwidth]{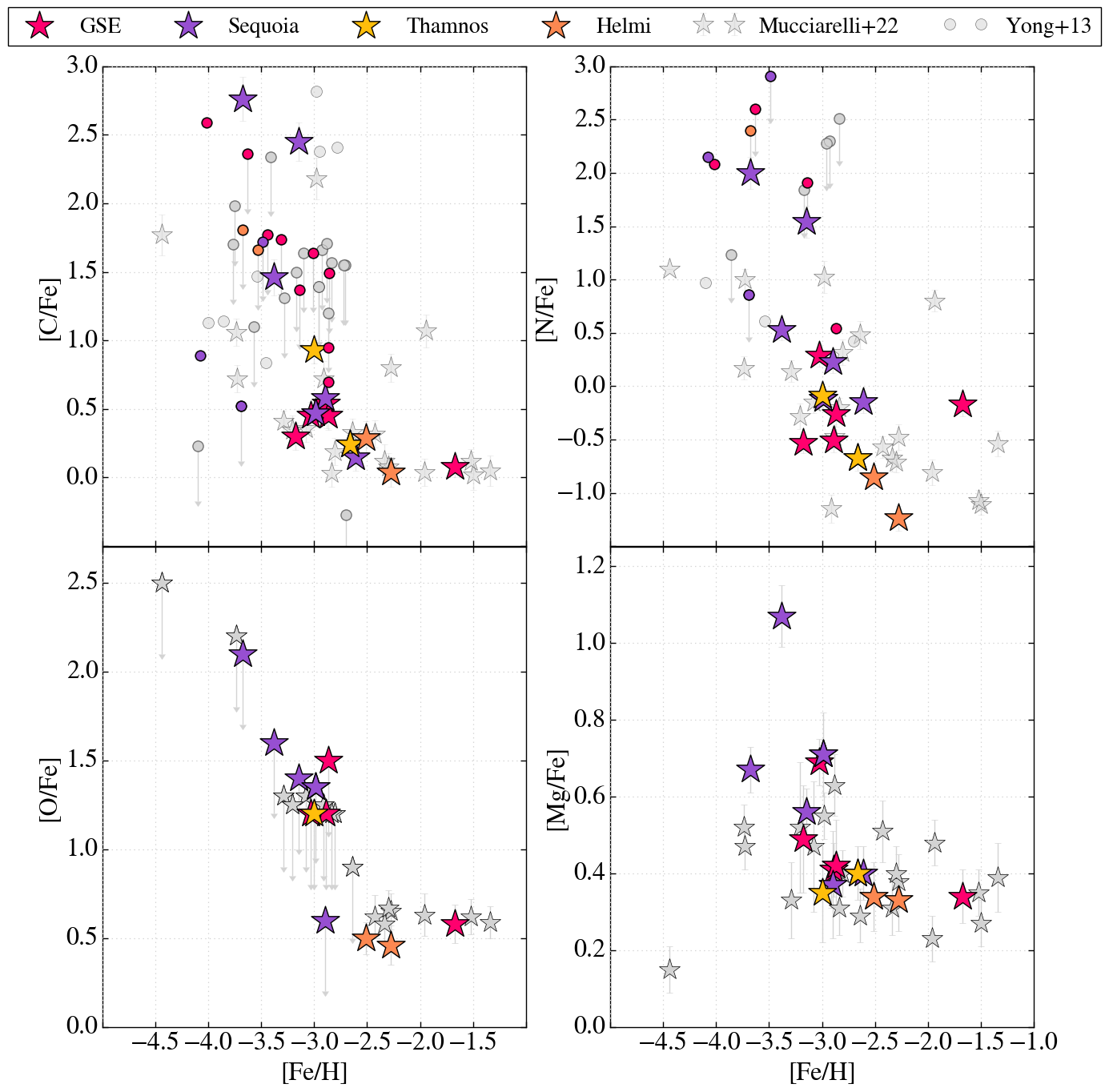}
    \caption{[C/Fe], [N/Fe], [O/Fe], and [Mg/Fe] vs [Fe/H] abundances for our sample stars. The colour code is the same as that for the boxes in Fig.\ref{fig:orbital para}. Unassociated stars are plotted in gray.}
    \label{fig:chem members}

\end{figure*}


Abundances of N, O, and Mg were obtained by a $\chi^2$-minimization between observed features and synthetic spectra computed with the code SYNTHE \citep{Kurucz2005} and adopting plane-parallel, 1D model atmospheres calculated with ATLAS9 \citep{Kurucz1993,Kurucz2005}. N abundances were derived by fitting the $A^{3}\Pi$-$X^{3}-\Sigma$ NH molecular band at 336 nm 
and adopting the linelist by \citet{fernando18}. O abundances were derived from the forbidden line at 6300.3 \AA \ for 10 stars. For other stars with spectra covering the O line region a detection was not possible, so we only give upper limits to the [O/Fe] ratios. Finally, Mg abundances were derived from the Mg b triplet or from the transition at 5528 \AA, depending on the metallicity of the star.

Uncertainties were estimated including the errors arising from the parameters and those from the fitting procedure. The uncertainty arising from the fitting procedure was obtained through Monte Carlo simulations, by creating for each star a sample of 500 synthetic spectra with the appropriate instrumental resolution and pixel-step and adding Poissonian noise to reproduce the observed signal-to-noise ratio (S/N). The line-fitting procedure has been repeated for these samples of simulated spectra, adopting as 1$\sigma$ uncertainty the dispersion of the  derived abundance distributions.

In order to increase the number of low-metallicity stars with measures of N, we considered also the targets by \citet{Yong13} 
located before the RGB Bump and, therefore, unmixed as those by \citet{Mucciarelli22}. We checked possible systematics between the two analyses. \citet{Yong13} adopted solar reference values by \citet{Asplund+09}, so we re-scaled their abundances to our 
solar abundances (see next paragraph). Both analyses adopted ATLAS9 model atmospheres. Finally, we checked that no significant offsets arise from the adopted CH and NH linelists. For CH, \citet{Yong13} adopted a first version of the linelist
by \citet{Masseron14}, also used by \citet{Mucciarelli22}. For NH, \citet{Yong13} used the Kurucz-gf reduced by a factor 
of two; we checked that this linelist provides N abundances fully compatible with those obtained with the linelist by \citet{fernando18}.

The stellar parameters and abundances for the sample of unmixed halo stars adopted in this paper, which includes new homogeneous N and O determinations for the sample of stars presented in \citet{Mucciarelli22}, are listed in Table~\ref{tab spectro}. Throughout this paper, the reference solar abundances are from \cite{gs98} but for O, for which we adopt the photospheric solar abundance determined by \citet{caffau11sun} from a CO$^5$BOLD 3D model of the solar atmosphere.

\subsubsection{Abundance scatter in CNO elements}

In Fig.\ref{fig1:observed scatter}, we present the measured [X/Fe] abundance ratios (with X = C, N, and O) as a function of [Fe/H] for the sample of unmixed stars of \citet[][star symbols]{Mucciarelli22}. The coloured boxes in each panel represent the observed spread in [X/Fe], defined as the difference between the maximum and the minimum abundance ratio, in metallicity bins of 1 dex.
The solid line shows the run of the mean value in the same metallicity bins. 

Figure~\ref{fig1:observed scatter} reveals a clear trend of increasing mean abundance ratio with decreasing metallicity for all elements (though it should be noticed that all [O/Fe] ratios for [Fe/H]~$< -2.5$ are upper limits). At the same time, towards lower metallicities the scatter of measured [X/Fe] abundances, $\Delta \rm [X/Fe]$, increases. For $\rm [Fe/H] > -1.5$, the spread is minimal, with $\Delta \rm [X/Fe] \approx 0$ ($\Delta \rm [X/Fe] \approx 0.6$) for [C/Fe] ([N/Fe]), but only 2 stars fall in this metallicity range.  However, in the range $-3.5 < \rm [Fe/H] < -2.5$, the scatter increases significantly to $\Delta \rm [X/Fe] \approx 1$ for [C/Fe] and $\Delta \rm [X/Fe] \approx 2$ for [N/Fe]. The observed scatter then apparently decreases for $\rm [Fe/H] <-4$, though it should be emphasized that the sample consists of only four stars within this range. Indeed, adding a sub-sample of data for low-metallicity stars from \citet[][circles]{Yong13} leads to larger scatters in both [C/Fe] and [N/Fe] (O measurements are not provided by \citeauthor{Yong13}). Expanding the sample size to include more  {homogeneous} abundance determinations of unevolved halo stars would allow a better estimate of the dispersion and it is, hence, highly desirable. 

\subsubsection{Orbital parameters}
\label{orbital-properties}


Our stellar sample does not include any kinematic cut, so within the metallicity range covered by the observations ($\rm -4.5 \le [Fe/H] < -1.0$ dex), we expect a mixture of stars with both in situ and accreted origin. To explore whether the origin of the abundance scatter is intrinsic or due to overlapping distributions of stars born in different environments, we study the dynamical properties of the stars in our sample.

It has been demonstrated that stars with common origin should remain clumped in the spaces defined by the integrals of motion (IoM, e.g., energy and angular momentum) even after they are completely phase-mixed within our Galaxy \citep{helmidezeeuw00} and that chemistry plays a crucial role in complementing the complicated dynamical information \citep[e.g.,][]{matsuno2022,horta2023,ceccarelli2024}. We thus exploit the extremely precise Gaia DR3 \citep{GaiaDR3} astrometry to get the 5D phase space information (i.e., position, proper motion, and parallax) for the stars in our sample. The spectroscopic line-of-sight velocities ($V_{\mathrm{los}}$) are taken from \citet{Norris13} and \citet{Mucciarelli22}. To reconstruct each star's orbit, we use the code \texttt{AGAMA} \citep{vasiliev19} assuming a \citet{mcmillan17} potential for the MW. The orbits are integrated in a reference frame with the Sun positioned $R_{\odot} = 8.122$~kpc away from the Galactic Center \citep{gravitycollaboration18} and $z_{\odot} = 20.8$ pc above the Galactic plane \citep{bennetandbovy19}. The three components of the Solar velocity are set to ($U_{\odot}$, $V_{\odot}$, $W_{\odot}$) = (12.9, 245.6, 7.78) km $\mathrm{s^{-1}}$ \citep{drimmelandpoggio18}, as computed by joining the proper motion of Sgr $\mathrm{A^{\ast}}$ \citep{reidandbrunthaler04} and the local standard of rest \citep{schonrich2010}. To estimate uncertainties on the orbital parameters, we run 10000 Monte Carlo simulations of the orbit for each star assuming Gaussian distributions for the uncertainties in the 6D phase space parameters. The uncertainties on the orbital parameters are defined as the 16th and 84th percentiles of their final distributions. The orbital parameters for the sample of unmixed halo stars adopted in this paper are listed in Table~\ref{tab orbits}.

In Fig.~\ref{fig:orbital para} we present the distribution of the IoM of our sample stars, alongside the substructures identified in the stellar halo \citep{dodd23}. As a reference, we plot in gray the IoM of stars from the APOGEE sample, using the Gaia values listed in the APOGEE DR17 catalogue \citep{Apogee} to compute the orbits. Given the limited statistics of our stellar sample, we opt to associate stars to a given substructure by looking directly at their position in the 3D IoM space (i.e., $E$, $L_{\mathrm{z}}$, and $L_{\perp}$). According to literature works based on clustering analyses on larger data sets \citep{ruiz-lara22,dodd23}, we define selection boxes in regions of the IoM spaces where different merging events are expected to have deposited their debris (see Fig.~\ref{fig:orbital para}). Among all the targets analysed in this work, we find 14 stars tentatively associated to the Gaia-Sausage-Enceladus (GSE) merging event \citep[pink box,][]{belokurov2018,helmi2018}, 8 from the Sequoia dwarf galaxy \citep[purple box,][]{myeong19}, 2 possibly related to Thamnos \citep[yellow box,][]{koppelman19} and, finally, 4 associated to the Helmi Streams \citep[orange box,][]{helmi99}. We find no matching stars with the ED2, ED3, Antaeus, and Sagittarius substructures \citep{ibata94,oria2022,dodd23}.

\begin{table*}
    \caption{Orbital parameters (orbital energy, angular momentum along the z-axis, perpendicular component of the angular momentum, and line-of-sight velocity) with uncertainties for the unmixed halo stars. }
    \centering
    \begin{tabular}{ c c c c c c} 
         \hline \hline 
\noalign{\vskip 0.2cm}
ID & ID Gaia DR3 & $E\pm \sigma(E)$  & $L_z \pm \sigma(L_z)$ & $L_\perp \pm \sigma(L_\perp)$ & $V_{los} \pm \sigma(V_{los})$ \\
&           & (10$^5$ km$^2$ s~$^{-2}$) & (10$^3$ kpc km s~$^{-1}$) & (10$^3$ kpc km s~$^{-1}$) &  (km s~$^{-1}$)  \\
  \hline
      \noalign{\vskip 0.1cm}

   \multicolumn{6}{c}{\citeauthor{Mucciarelli22}'s (\citeyear{Mucciarelli22}) sub-sample}\\
    \hline
        \noalign{\vskip 0.1cm}

     BD-012582 & 3602201479317188608 &              $-$1.7146$\pm$ 0.0006 &             0.527 $\pm$0.010 &                0.840 $\pm$0.006 &            1.28 $\pm$0.10 \\
  CD-241782 & 5085629204207431680 &              $-$1.7365 $\pm$0.0033 &            $-$0.800 $\pm$0.021 &                0.087 $\pm$0.004 &          101.39 $\pm$0.10 \\
   CD-30298 & 5031398163988267008 &              $-$1.3808 $\pm$0.0076 &             1.437 $\pm$0.008 &                0.142 $\pm$0.003 &           27.63 $\pm$0.10 \\
CS22183-031 & 2482988183718639616 &              $-$1.5933 $\pm$0.0241 &            $-$0.296$\pm$0.142 &                0.144$\pm$0.004 &           16.73$\pm$0.03 \\
CS22186-023 & 4868824473488624640 &              $-$1.6332 $\pm$0.0023 &             1.123 $\pm$0.024 &                0.638 $\pm$0.035 &           52.05$\pm$0.06 \\

\ldots & \ldots & \ldots & \ldots & \ldots & \ldots\\
  \hline
      \noalign{\vskip 0.1cm}

    \multicolumn{6}{c}{\citeauthor{Yong13}'s (\citeyear{Yong13}) sub-sample}\\
    \hline
    \noalign{\vskip 0.1cm}
 52972-1213-507 &  811812182598263552 & $-$1.4841 $\pm$ 0.0369 & $-$0.78 $\pm$ 0.286 &   1.47 $\pm$ 0.048 & $-$177.00 $\pm$ 0.50 \\
53327-2044-515 &  290930261314166528 & $-$1.6383 $\pm$ 0.0006 &  0.80 $\pm$ 0.013 &   0.39 $\pm$ 0.024 & $-$193.50 $\pm$ 0.50 \\
53436-1996-093 &  767103875148269696 & $-$1.3353 $\pm$ 0.0610 &  0.88 $\pm$ 0.110 &   1.54 $\pm$ 0.190 &  $-$15.30 $\pm$ 1.70 \\
54142-2667-094 &  598558225898219776 & $-$1.6346 $\pm$ 0.0120 &  0.64 $\pm$ 0.098 &   1.15 $\pm$ 0.108 &   43.40 $\pm$ 0.70 \\
  BS-16545-089 &  760839819965612928 & $-$1.4938 $\pm$ 0.0098 &  0.34 $\pm$ 0.065 &   0.42 $\pm$ 0.038 & $-$161.10 $\pm$ 0.50 \\
\ldots & \ldots & \ldots & \ldots & \ldots & \ldots  \\
  \hline
    \end{tabular}
    \tablefoot{The table is available in its entirety at the CDS. A portion is shown here for guidance regarding its form and contents.
    }
    \label{tab orbits}
\end{table*}

As shown in Fig.~\ref{fig:chem members}, the abundance ratios of different substructures tend to overlap in the [C/Fe], [N/Fe], [O/Fe], and [Mg/Fe] versus [Fe/H] diagrams\footnote{ {Given the paucity of O abundance determinations for [Fe/H]~$< -$2.5, we opt to include Mg to highlight the presence of a scatter in the $\alpha$-element abundances. The early evolution of elements other than CNO ones will be discussed in a forthcoming paper.}}. No obvious distinct trends are observed, even though marginal evidence that some Sequoia stars are among the most enhanced in all of the elements exist. Larger samples are clearly required to derive a robust conclusion; still, with the data at hand we can hazard an interpretation of the spread in the elemental abundances as not due to the sum of distinct but tighter patterns, related to individual mergers. The observed scatter at these very low metallicities seems to be due to some intrinsic processes.

\subsection{High-redshift systems}
\label{samplehigh}

Recent JWST observations have captured rest-frame near-ultraviolet (UV) and optical spectra of distant galaxies exhibiting emission lines of C, N, and O from which accurate abundances relative to H have been determined \citep[e.g.,][]{arellano2022,schaerer2022,heintz2023,Cameron23}. While most spectra obtained thus far closely resemble those of typical metal-poor star-forming galaxies in the local universe, a few luminous, low-metallicity systems display  {anomalous super-solar C/O or N/O abundance ratios \citep[e.g.,][]{Cameron23,deugenio2024,MarquesChaves2024}}. This has given rise to various speculations about the origin of such peculiar chemical features. It has been suggested that we might be either witnessing the formation of proto-globular clusters, or detecting the signatures of Wolf-Rayet or AGB stars that lose their outer layers through stellar winds, or having a glimpse into early ISM pollution from Pop~III and/or very massive stars ($m_\star \ge 1000$~M$_\odot$) forming from collisions in dense clusters, or even seeing the effects of the activity of an active galactic nucleus \citep[see][and references therein]{MarquesChaves2024}.

In Sect.~\ref{reshighz}, we develop GCE models tailored to reproduce the observed properties of the prototype  {extreme C- (GS-z12) and N-emitters (GN-z11),} resting on our benchmark model for the MW. We compare the predicted abundances with CNO abundance estimates for  {GS-z12,} GN-z11, and other similar systems \citep{Bunker23,Cameron23,Isobe23,Maiolino23,deugenio2024,Harikane24,Ji_GN-z11,Senchyna2024,Xu24}.

\section{Chemical evolution models}
\label{models}

The increasing star-to-star scatter in chemical abundance ratios at low metallicity serves as a powerful tool to understand the earliest phases of the MW chemical enrichment \citep{Argast2004, Cescutti2015, Wehmeyer2015, Griffith23, Vanni23}. To gain insights into the early phases of CNO element evolution, we employ a GCE model that has been extensively tested against CNO abundance data of our and external galaxies \citep{Romano17,Romano19,Romano2020}. While the main focus of previous work was on [Fe/H]~$> -1.5$ environments, here we add a detailed treatment of the first, inhomogeneous phases of galactic chemical enrichment (Sect.~\ref{model implementations}).

\subsection{The Milky Way model}
\label{MW model}

Hereafter, we provide an overview of the principal features, assumptions, and limitations of the adopted GCE model for the MW, which is constantly developed and improved. Further details are available in the quoted papers.

The model utilizes a multizone framework, dividing the Galactic disc into concentric annuli of 2 kpc each. It assumes that the in-situ inner halo and thick-disc components form from an initial phase of accretion of matter of primordial chemical composition \citep{Chiappini97,Chiappini01}. During the first $\sim$3--4~Gyr of evolution \citep[see][for the calibration of the model against asteroseismic stellar ages]{Spitoni2019,Spitoni2021}, an efficient burst of star formation results in rapid gas depletion, leading to the formation of the most ancient stellar populations. The thin disc forms subsequently from a second, nearly independent infall event, on time scales ranging from $\sim 3 \: \rm Gyr$ for the inner disc up to a Hubble time for the outer regions \citep{Romano20,Chiappini01}. This `inside-out' formation of the Galactic disc is required to reproduce the abundance gradients and the gas distribution along the disc \citep[see][and references therein]{Matteucci1989}.

The star-formation rate is described by the Schmidt-Kennicutt law \citep{Kennicutt98}, i.e.,  $\psi (R, t) = \nu \, \Sigma_{gas}(R, t)^k$. Here, the star-formation rate is directly proportional to the gas surface density, $\Sigma_{\text{gas}}$, $k$ is set equal to 1.5, and the star-formation efficiency, $\nu$, is constant along the disc during the halo/thick-disc phase, while it becomes a function of the distance from the Galactic center during the thin disc formation \citep[see][]{Palla20}.
 The newly formed stellar mass is distributed in the mass range 0.1--100~ M$_{\odot}$ according to the \cite{Kroupa1993} IMF, characterized by a slope of 1.7 in the high-mass regime.

\subsection{Nucleosynthesis prescriptions}
\label{chem prescriptions}

The chemical enrichment process is followed by tracking the evolution of all stable chemical species from hydrogen (H) to europium (Eu) and by considering the mass-dependent stellar evolutionary timescales, i.e., relaxing the instantaneous recycling approximation. This enables to adequately follow the evolution of chemical elements that are produced on different time-scales by stars of different initial mass and chemical composition.

For Pop~II stars, we employ stellar yields dependent on the initial mass and metallicity of the stars that allow a satisfactory fit to CNO abundance data for different MW components and early-type galaxies \citep[][see also Romano et al. in prep]{Romano19,Romano2020}. In particular, for low- and intermediate-mass stars (LIMS) with $1 \le m_{\star}/\rm{M}_{\odot} \le 6$, we retrieve the stellar yields for non-rotating stars from the FRUITY database \citep{Cristallo2009,Cristallo2011,Cristallo2015}. For massive stars ($13 \le m_{\star}/\rm{M}_{\odot} \le 120$) that explode as CCSNe, we use the yields of \citet[][their fiducial set R]{Limongi+18} computed for initial metallicities [Fe/H]~= $-3$, $-2$, $-1$, and 0 and accounting for different rotational velocities (0, 150, and 300~km~s$^{-1}$). For type Ia SNe (SNeIa, exploding white dwarfs in binary systems), we assume the stellar yields of \cite{Iwamoto99}. SNIa primarily contribute to the bulk of iron production during later evolutionary stages and play a significant role in the decline of the $\rm [\alpha/Fe]$ ratio at metallicities higher than investigated here. The nova outburst rate is incorporated following the methodology outlined in \cite[][and references therein]{Romano99,Romano21}. No contribution from novae is expected before 1~Gyr has elapsed from the beginning of star formation.

\subsection{High-redshift galaxies: GN-z11 and GS-z12}
\label{high redshift galaxy}

As part of this study, we aim to reproduce  {the peculiar CNO abundances of a few young systems observed at high redshift. GN-z11 is a compact object} with half-light radius $\le$ 200~pc observed at $z \simeq 10.6$, hosting a $10^{8 \sim 9.1}$~M$_\odot$ stellar population and characterized by a low-metallicity, N-rich ISM, 12+log(O/H) = $7.84^{+0.06}_{-0.05}$, N/O~$\simeq 4 \times$ solar \citep{Bunker23, Cameron23, Maiolino23, Tacchella23, Senchyna2024}.  {\citet{Cameron23} set a lower limit of log(C/O)~$> -0.78$ to its carbon-to-oxygen ratio, yet we note that other extreme N-emitters found at high redshift have C/O ratios fairly similar to the ones of local metal-poor galaxies \citep{MarquesChaves2024,Schaerer2024}. The galaxy GS-z12 ($z \sim 12.5$), with stellar and gaseous masses of $\sim5 \times 10^7$~M$_\odot$ and $\sim10^7$~M$_\odot$, respectively, and typical size of 100~pc has instead a carbon content unexpectedly high for its low metallicity, log(C/O)~$> -0.21$ at log(O/H)+12~= $7.59 \pm 0.25$ \citep{deugenio2024}.}

To model these high-redshift galaxies, we run  {single-zone GCE models. Alike for the MW, fresh gas is accreted according to an exponentially decreasing law, $\mathrm{d}M_\mathrm{inf}/\mathrm{d}t \propto \mathrm{e}^{-t/\tau}$, where $M_\mathrm{inf}$ is the total mass accreted and $\tau$ is the infall time scale. We assume that metal-free gas is quickly accreted ($\tau \sim$ 1--3~Myr) and efficiently converted into stars ($\nu \sim$ 1--10 Gyr$^{-1}$)} within compact systems characterized
by surface gas densities much higher than that of the MW  {(at the end of the simulation for GN-z11, the total surface mass density is $\Sigma \simeq 7.5 \times 10^3$~M$_\odot$ pc$^{-2}$, to be compared to values more than two orders of magnitude lower in the solar neighbourhood)}. The gas is converted into stars according to the Schmidt-Kennicutt law and the IMF is taken from \citet{Kroupa1993}, similarly to what is
done for the MW (see Sect.~\ref{MW model}). The adopted nucleosynthesis prescriptions for Pop II stars are those described in Sect.~\ref{chem prescriptions}.
The free parameters of the model  {are fixed as to reproduce the observed properties of GN-z11 and GS-z12. Specifically, in the model for GN-z11 we assume the star formation history of \citet{Tacchella23}, namely, a low-level star formation episode ($\nu =$~1~Gyr$^{-1}$) lasting 40~Myr, followed by a burst ($\nu =$~10~Gyr$^{-1}$) of longer duration, 60~Myr, that forms most of the stellar mass of the system. This vigorous burst is fed by prominent infall of gas of primordial chemical composition that gives birth to a second generation of Pop~III stars, overwhelming the contribution to the chemical enrichment from previous generations of Pop~II stars (see Sect.~\ref{gnz11}). In the model for GS-z12, a shorter (10~Myr long) but similarly efficient ($\nu \sim$ 10 Gyr$^{-1}$) star-formation burst is considered, resulting in a constant star formation rate of 1.6~M$_\odot$ yr$^{-1}$ and stellar mass at the time of the observations (350~Myr after the Big Bang) of log($M_\star/M_\odot$)~= 7.60, in agreement with the values of $1.62^{+0.29}_{-0.24}$~M$_\odot$ yr$^{-1}$ and log($M_\star/M_\odot$)~= $7.68 \pm 0.19$ inferred by \citet{deugenio2024}. Finally, to match the low metallicity observed for GN-z11 in spite of its powerful starburst, we have to assume that a galactic outflow vents out more than a half of the stellar ejecta. This seems a reasonable request, since massive galactic outflows are actually known to be a key mechanism to regulate, or even quench, the star formation of high-redshift starbursts \citep[e.g., ][and references therein]{jones2019}. No galactic outflows are required for the smaller galaxy, GS-z12.}

\section{Model implementation}
\label{model implementations}

\begin{figure*}
\centering
\includegraphics[width=\textwidth]{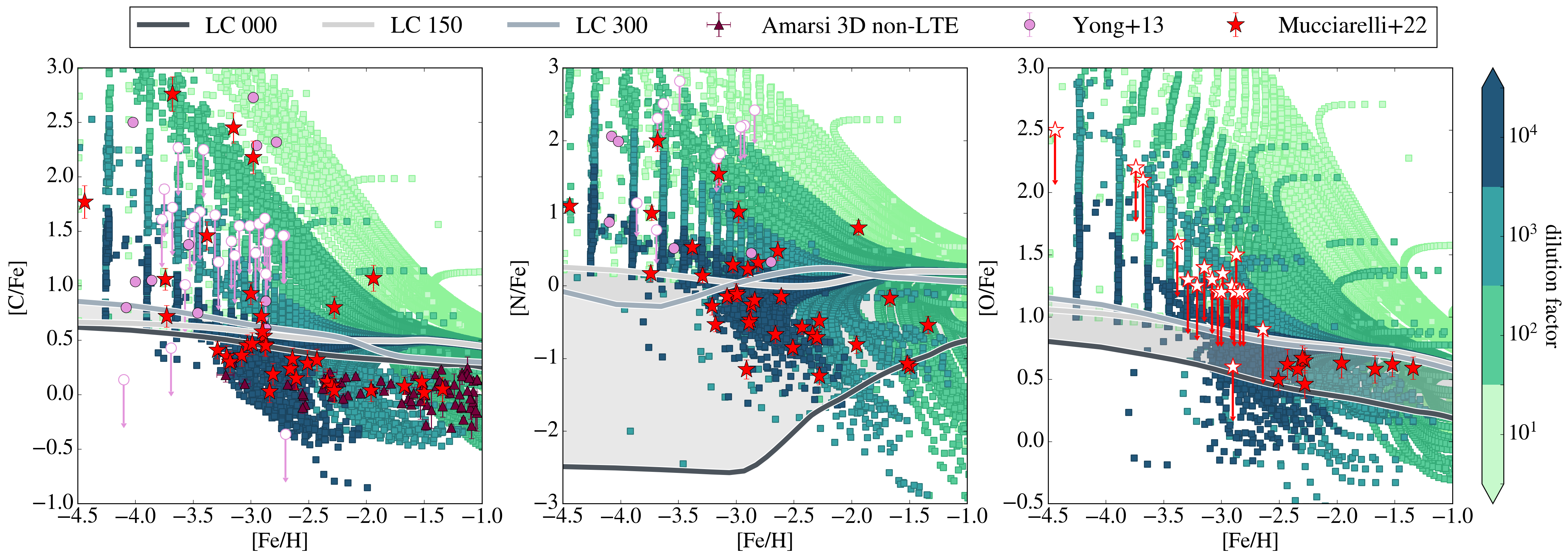}
    \caption{[C/Fe] (left), [N/Fe] (middle) and [O/Fe] (right) vs [Fe/H] abundances predicted by our stochastic chemical evolution model for the Galactic halo (squares). Different shades of green represent different dilution factors \citep[see text and][]{Nandal24Massive, Nandal24PopIII}. Red stars, purple circles, and brown triangles represent, respectively, the data sets of \cite{Mucciarelli22}, \cite{Yong13}, and \cite{Amarsi19}. Empty symbols represent upper limits. The continuous curves in shades of grey represent the predictions of the homogeneous GCE model without pre-enrichment from Pop~III stars, assuming the yield sets of \cite{Limongi+18} with different rotation velocities for massive stars.}
    \label{fig:1}

\end{figure*}

The role of Pop~III stars is frequently overlooked in chemical evolution models. However, while at high metallicity ([Fe/H]~$> -$1) stars are expected to form from a thoroughly mixed medium enriched by multiple stellar populations, which effectively washes out the distinctive chemical signatures of Pop~III stars, at lower metallicities the contribution from the first massive stars becomes fundamental to explain the peculiar chemical abundances of ancient, metal-poor stars that survived to the present day \citep[][]{ioanna2023, Vanni23, Rossi2024arXiv}. On the other hand, during the initial phases of galaxy evolution, stochastic processes associated with the star formation itself become crucial in shaping the evolution and chemical enrichment history of galaxies \citep[e.g.,][]{Caplar19,Tacchella20, Wang22, Pallottini23,Iyer24}.

To model the early chemical enrichment phases of our Galaxy we implement the GCE model described in Sect.~\ref{models} in order to account for a pre-enrichment driven by Pop~III stars and a stochastic star formation process. We assume that in the initial phases of formation, the MW is built up by gas clumps with different gas densities that merge into the MW main branch. We assume that a single Pop~III SN precursor is formed in each clump, with properties (mass, internal mixing level, and explosion energy) randomly selected from the \cite{Heger+woosley10} model grid. The mass of Pop~III stars is stochastically selected within the mass range $10 \le m_{\rm Pop\: III}/{\rm M}_{\odot} \le 100$, while the explosion energy, $\rm E_{SN}$, is assigned by randomly choosing among four different types of supernovae, each with a specific energy range: faint ($\rm E_{51}  $\footnote{ $\rm E_{51}= E_{SN}/10^{51}$.}= 0.3--0.6~erg), core-collapse  ($\rm E_{51} =$ 1.2--1.5 erg), high-energy ($\rm E_{51} =$ 1.8--3.0~erg), and hypernovae ($\rm E_{51} =$ 5.0--10 erg).
We have not included the pre-enrichment from PISNe in our analysis. In fact PISNe, with initial masses ranging from 140 to 260 M$_{\odot}$, are predicted to enrich the surrounding ISM up to $\rm [Fe/H] \approx -2$ while underproducing N, $\rm [N/Fe] < -2 $ \citep{Salvadori2019}, and no stars with such low N abundances are present in our sample (see Sect.~\ref{sample}). The clumps are assumed to be characterized by different dilution factors \citep[$10^1, 10^2, 10^3, 10^4$, see also][]{Nandal24Massive, Nandal24PopIII}, defined as the ratio between the hydrogen mass and the mass ejected by Pop~III SNe. This factor regulates how much the SN ejecta is diluted by the pristine surrounding medium and is a free parameter of our model. Note that assuming different dilution factors is equivalent to consider clumps of different sizes/densities.
Finally, for each dilution factor, we consider 200 realisations of the chemical enrichment. 

After the explosion of Pop~III SNe, the metallicity of the gas in each clump reaches a level that triggers the formation of Pop~II stars. We then track the subsequent chemical evolution assuming the yields of \cite{Limongi+18} for massive stars in the case of fast rotators with $v_{\rm rot} = 150$~km~s$^{-1} $ and the FRUITY yields for LIMS (see Sect.~\ref{chem prescriptions}).

\section{Early chemical enrichment: exploiting long-lived stars in the Milky Way}
\label{resMW}

In this section, we present the results of chemical evolution models for the MW. In Fig.~\ref{fig:1} we show the predicted abundance ratios of CNO elements as a function of [Fe/H] compared with observational data \citep{Mucciarelli22, Yong13, Amarsi19}. Different shades of green represent the predictions of the stochastic GCE model obtained assuming different dilution factors for the Pop~III star ejecta, i.e., different sizes/densities for the star-forming clumps at the beginning of the simulation (see Sect.~\ref{model implementations}). 
The first thing to note is that, to reproduce the full range of [Fe/H] values covered by our stellar sample (Sect.~\ref{samplehalo}), we need to account for different dilution factors\footnote{ {Hydrodynamical simulations of the expansion of SN bubbles suggest that different dilution factors have to be expected in dependence of the total energy, ejecta mass, and ISM densities \citep[e.g.,][and references therein]{Mori2002,Romano2019hydro}. A proper treatment of the ISM dynamics and energetic is still challenging \citep[e.g.,][]{Kim2023} and well beyond the scope of this work.}} ranging from $10^{1}$ to $10^{4}$. Specifically, the higher dilution factors allow us to reproduce the bulk of the observational data at $\rm [Fe/H] < -2 $. 
In this case, the ejecta from the first SNe are dispersed in denser (or larger) gas knots that contain higher amounts of neutral hydrogen, resulting in a shift of [Fe/H] towards the low metallicity regime. As the dilution factor decreases, the iron released by Pop~III SNe is diluted in a less dense environment, meaning the stellar clump is less compact, or in a smaller bubble, resulting in an increase in [Fe/H] abundance. Note that our assumptions on the dilution factors only affect the [Fe/H] abundances, which depend on the total amount of neutral hydrogen present in the stellar clumps, while the [X/Fe] ratios remain independent from these assumptions.

\begin{figure*}
\centering
\includegraphics[width=0.7\textwidth]{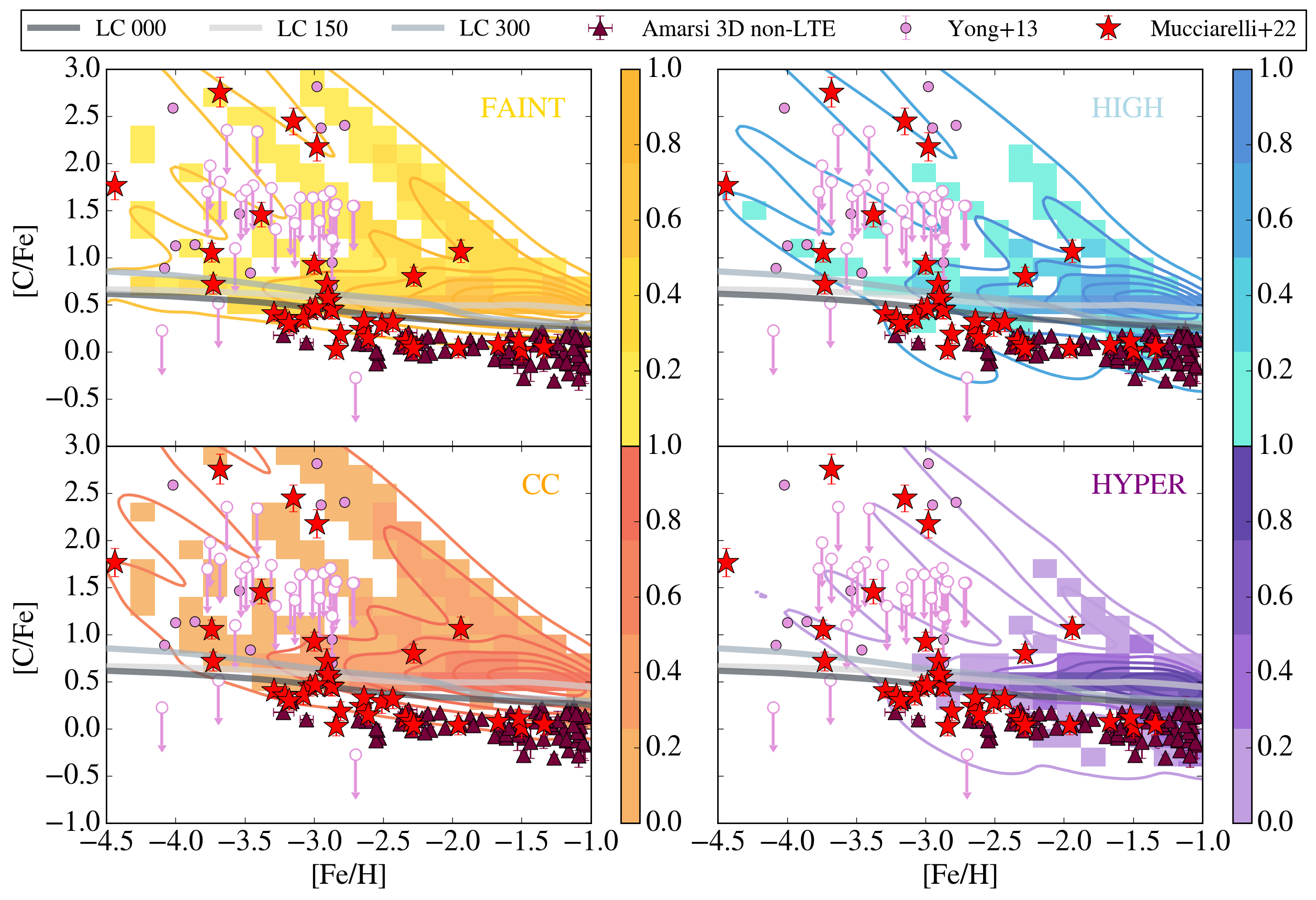}
    \caption{ Predicted density maps of [C/Fe] vs [Fe/H] for stellar populations inhabiting the MW halo. The different panels/colours highlight the contributions of various Pop~III SN types: faint, core-collapse, high-energy, and hypernovae. 
    The grey continuous lines represent the predictions of the homogeneous model without the contribution of Pop~III SNe, for different rotation velocities of Pop~II massive stars \citep[$v_{\rm rot} =$ 0, 150, and 300~km s$^{-1}$, ][]{Limongi+18}. Data (symbols, see figure legend) are from \cite{Mucciarelli22}, \cite{Yong13}, and \cite{Amarsi19}. Empty symbols represent upper limits.}
    \label{fig:2}

\end{figure*}

\begin{figure*}	
\centering
\includegraphics[width=0.7\textwidth]{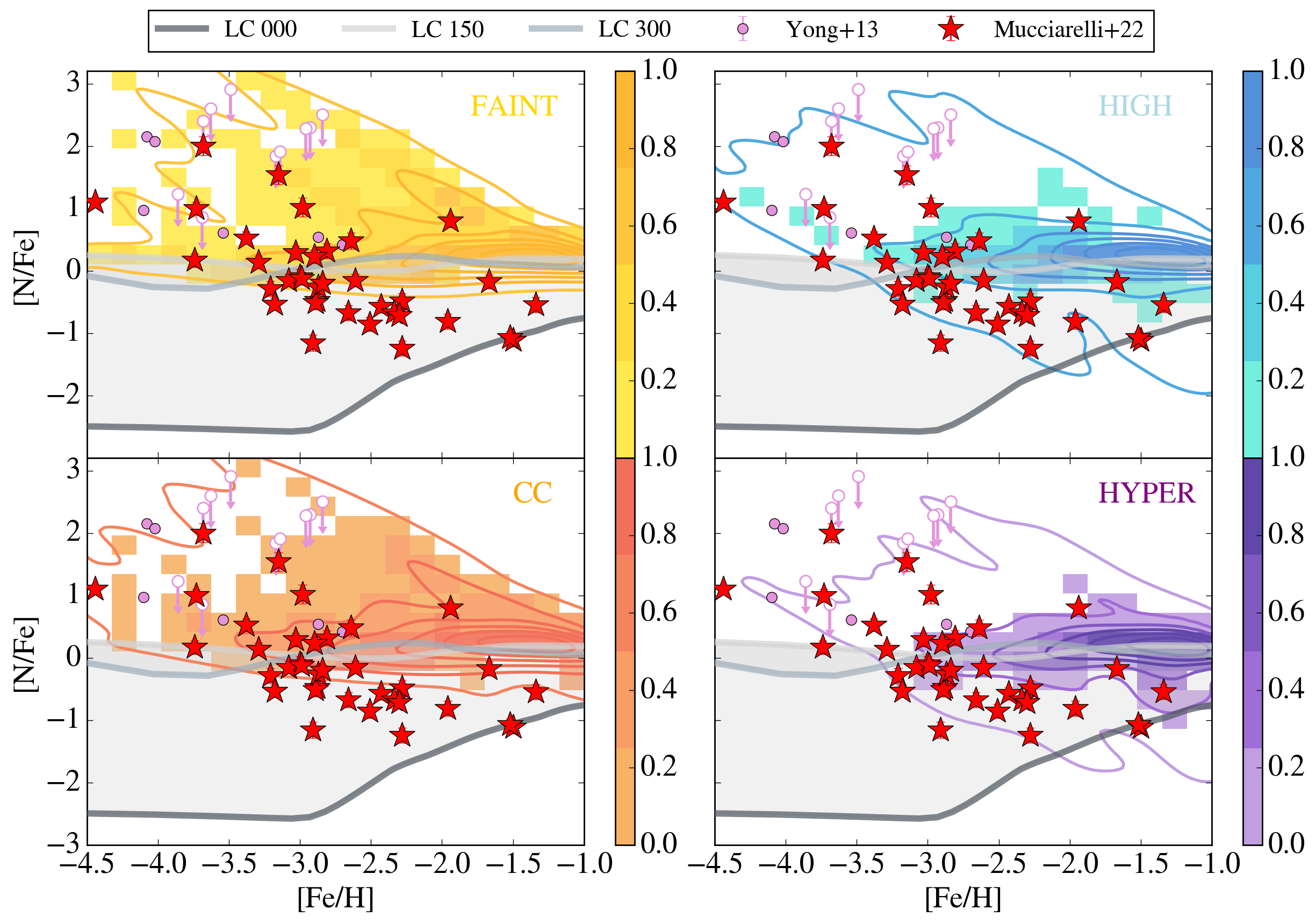}
  \caption{Same as Fig.~\ref{fig:2}, in the [N/Fe]--[Fe/H] space. Nitrogen abundances for the unmixed RGB stars of \cite{Mucciarelli22} have been derived as described in Sect.~\ref{sampleale22}.}
   \label{fig:3}
\end{figure*}

\begin{figure*}	
\centering
\includegraphics[width=0.7\textwidth]{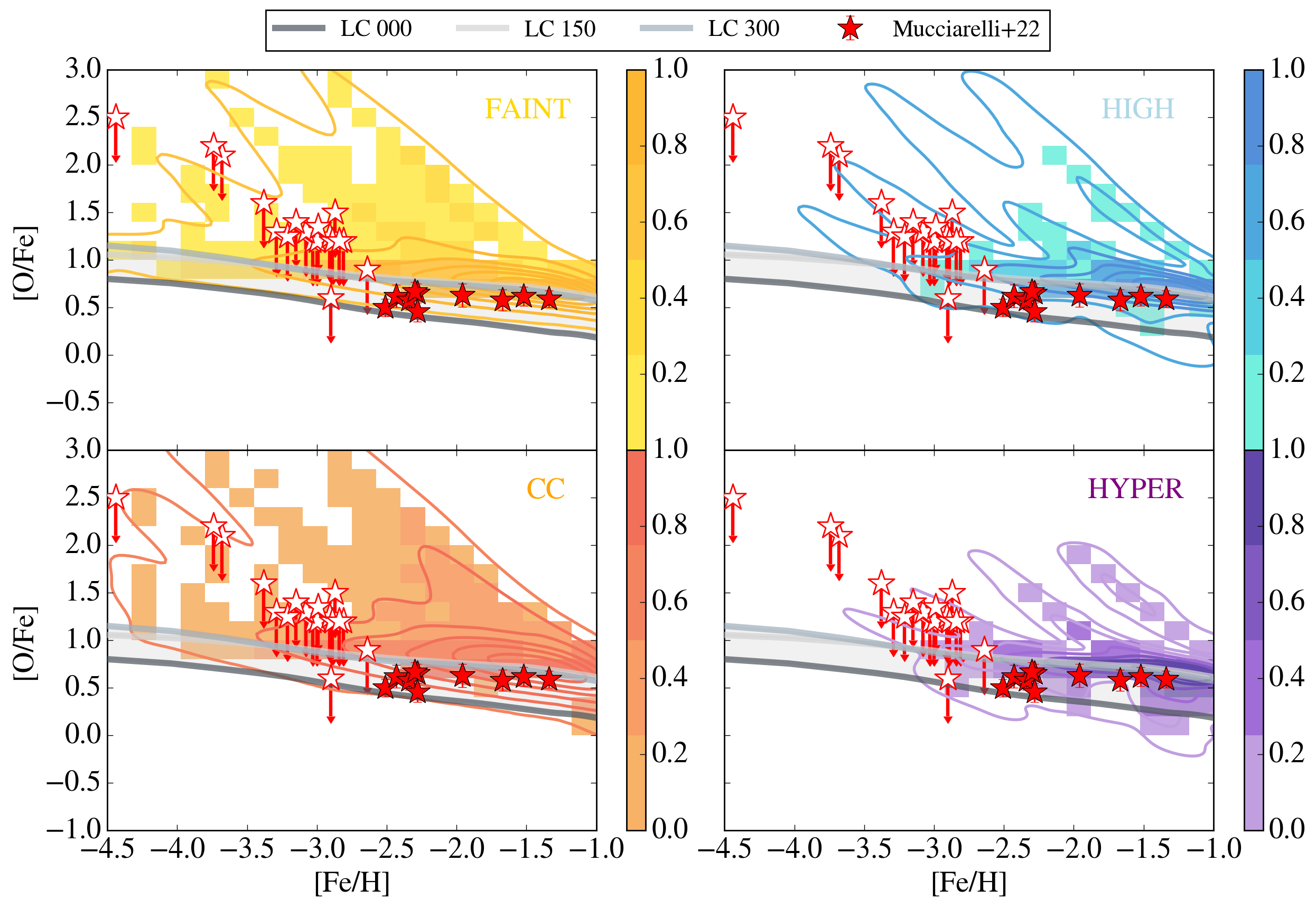}
  \caption{Same as Fig.~\ref{fig:3}, in the [O/Fe]--[Fe/H] space.}
   \label{fig:proboxigen }
\end{figure*}

\begin{figure*}
\begin{center}$
\begin{array}{cc} 
\includegraphics[width=0.7\textwidth]
{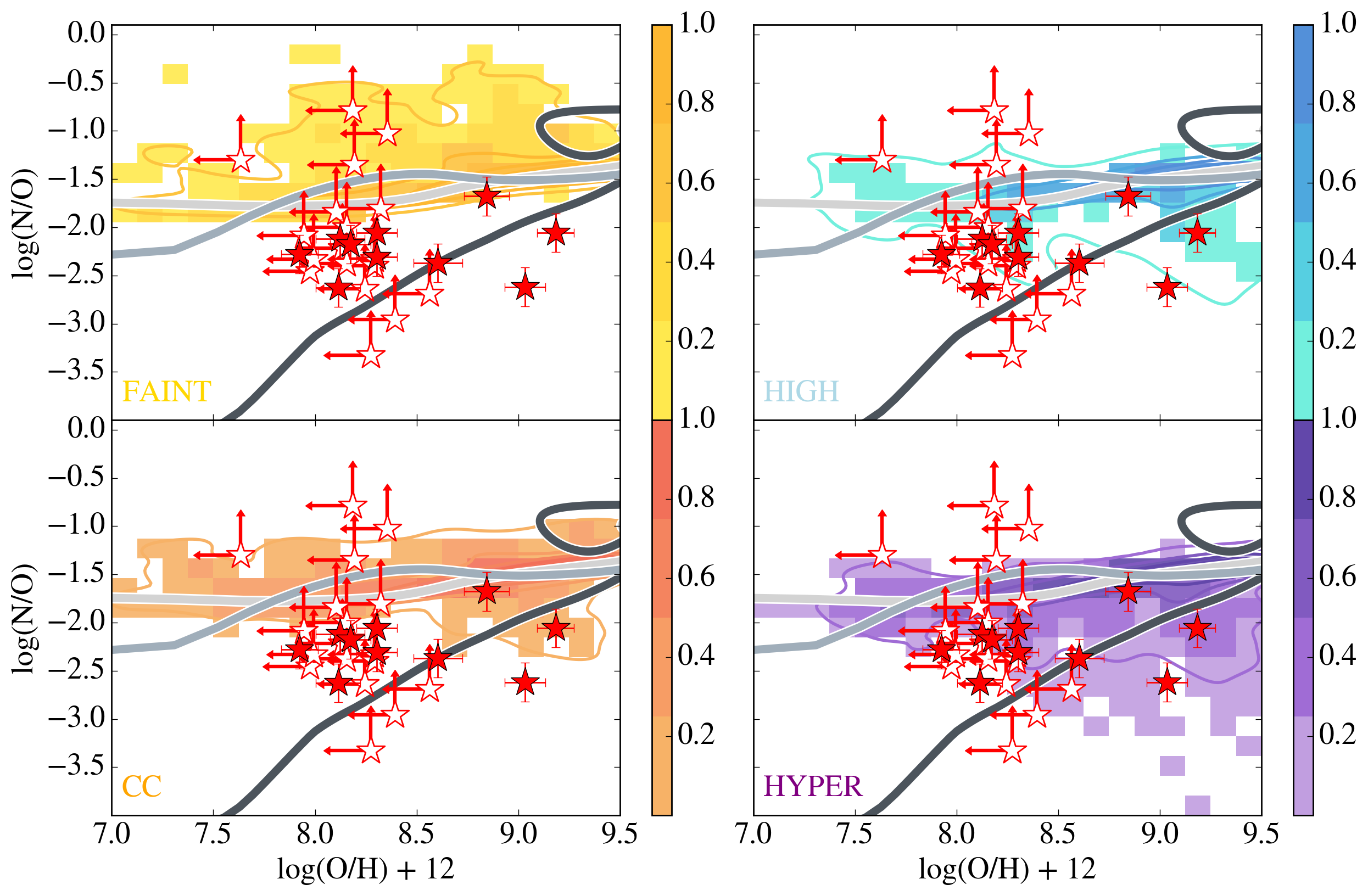} 
\end{array}$
\end{center}
   \caption{Same as Fig.~\ref{fig:3}, in the $\rm log(N/O)$--$\rm log(O/H) + 12$ space. Nitrogen and oxygen abundances for the unmixed RGB stars of \cite{Mucciarelli22} have been derived as described in Sect.~\ref{sampleale22}.}
   \label{fig:no_ prob}
\end{figure*}

\begin{figure*}
\begin{center}$
\begin{array}{cc} 
\includegraphics[width=0.7\textwidth]
{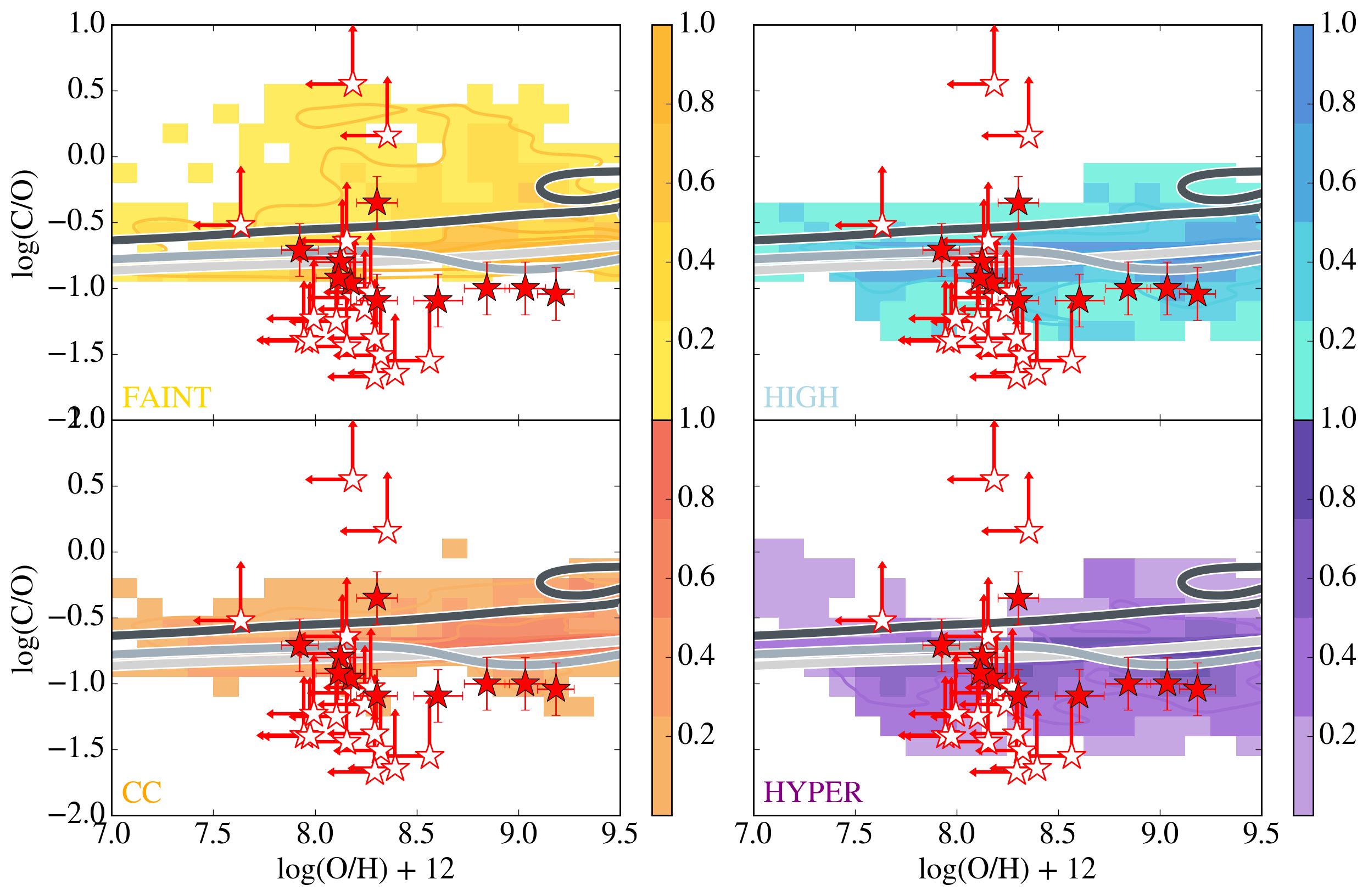}
\end{array}$
\end{center}
   \caption{Same as previous figure, but for carbon. Carbon and oxygen abundances for the unmixed RGB stars of \cite{Mucciarelli22} have been derived as described in Sect.~\ref{sampleale22}.}
   \label{fig:co_ prob}
\end{figure*}

As shown in Fig.~\ref{fig:1}, our stochastic GCE model successfully reproduces the scatter observed at low metallicity for both C, N, and O, even if it is worth noting that for $\rm [Fe/H]<-2.5$ we could only set upper limits to the O abundances. In Fig.~\ref{fig:1} we also compare our results with the predictions of homogeneous GCE models that do not incorporate pre-enrichment from Pop~III stars. The homogeneous GCE models adopt yields for Pop~II massive stars from \cite{Limongi+18}, computed with different initial rotational velocities ($v_{\rm rot} =$~ 0, 150, and 300~km~s$^{-1} $) and metallicities ([Fe/H]~= $-$3, $-$2, $-$1, and 0) of the stars. Notice that for [Fe/H]~$< -$3, we use the yield set corresponding to the lowest [Fe/H] value. 
It is clearly seen that, regardless of the assumed rotation velocities, homogeneous models fail to reproduce the observed [X/Fe] abundance ratios at $\rm [Fe/H]< -2.5$. This is particularly evident for C and N and it is, at least partly, due to the fact that the yield grid of \cite{Limongi+18} does not include $Z = 0$ stellar models. Indeed, at higher metallicities ($\rm [Fe/H] \gtrsim -2.5$) the models reach a satisfactory agreement with the observations. There is a clear tendency to overproduce C, though. Regarding the [N/Fe] abundance ratio, we notice that models assuming massive star yields computed with different rotational velocities adequately represent stars with low N abundances ($\rm [N/Fe]< 0$) across the full metallicity range. Nevertheless, even when assuming fast rotation, which boosts primary N production at low metallicities \citep{Meynet06,Limongi+18}, these models fall short of reproducing N-enriched stars.

\subsection{The origin of the scatter}
\label{different energies}

Having calibrated our stochastic GCE model to reproduce the dispersion of our data sample through an appropriate choice of the dilution factors, we now delve into the nature of the polluters responsible for the variation of chemical abundance ratios within our stellar sample at low metallicity.

 {In Figs.~\ref{fig:2} to \ref{fig:co_ prob},} we present `chemical density maps' for the present-day stellar population in the MW halo, as predicted by our model. In all figures, different panels refer to different explosion energies of Pop~III SNe contributing to the initial phases of chemical enrichment. As it is evident from the figures, different Pop~III SN types contribute differently to the observed scatter. Low-energy Pop~III SNe (faint and CCSNe) are responsible for high values of  {[C/Fe]~$> 1$, [N/Fe]~$>0.5$, and [O/Fe]~$> 1.5$} and struggle to accurately reproduce the regions with [C/Fe]~$< 0.5$, [N/Fe]~$<0$,  {and [O/Fe]~$< 1$}.  As the explosion energy of Pop~III SNe increases (high-energy SNe and hypernovae), the [C/Fe]~$< 0.5$  and  {[O/Fe]~$< 1$} data points are well reproduced. In contrast, despite accounting for the influence of energetic Pop~III SNe, regions at $\rm [N/Fe] < -0.5$ are sparsely recovered. This discrepancy does not necessarily indicate an issue, but rather suggests that the predominant polluters responsible for enriching these stars could be Pop~II stars with diverse rotation velocities. At this point, it is essential to clarify that our study does not aim to identify the potential descendants of the first stars. Although our results seem indeed to suggest that among  {carbon-, nitrogen-, and oxygen-enhanced stars} there may be possible candidates, it is beyond the scope of this work to engage in such speculation. The main goal of this paper is to shed light onto the early evolution of our own and other galaxies, accounting for the contribution of all possible stellar polluters acting at early times.

In the light of the comparison of our results for the MW with what can be observed at high redshift, it is important to discuss the density maps  {for $\rm log(N/O)$ and  $\rm log(C/O)$ vs $\rm log(O/H) + 12$.} These are presented  {in Figs.~\ref{fig:no_ prob} and \ref{fig:co_ prob},} although the presence of many upper/lower limits in our stellar sample prevents us from drawing definitive conclusions. As it is evident from the figures, our predictions cover a broad range in metallicity, $\rm 7<log(O/H)+12<9.5$. Notably, there is a clear trend in the abundance of $\rm log(N/O)$  {and  $\rm log(C/O)$} with varying energy levels of Pop~III SNe.
With pre-enrichment from faint SNe, it is possible to achieve extremely high values of $\rm log(N/O)$ up to $\approx -0.5$  {and  $\rm log(C/O)$ up to 0.5} . As the energy of Pop~III SNe increases, regions with $\rm log(N/O) > -1$  {and  $\rm log(C/O) > 0$}  become less populated, while regions with $ \rm -2.5 < log(N/O) < -1$  {and  $\rm -1 < log(C/O) < 0 $}  at $\rm 7.5 < log(O/H) < 9.5$ start to fill, reflecting the influence of core-collapse and high-energy SNe. Finally, with hypernovae, areas with higher metallicity, $\rm log(O/H) > 8.5$, and low values of $\rm log(N/O) < -2.5$  {and  $\rm log(C/O) < -1$}  are well reproduced.
It is important to highlight the effects of the two extreme Pop~III SN energetic regimes: only faint SNe manage to populate the regions with $\rm log(N/O) > -1$  {and  $\rm log(C/O) > 0$}, while hypernovae primarily impact regions with $\rm log(N/O) < -2.5$  {and  $\rm log(C/O) < -1$} .

\section{Early chemical enrichment: the gaseous medium of high-redshift galaxies}
\label{reshighz}


\begin{figure*}
\centering
\includegraphics[width=0.9\textwidth]{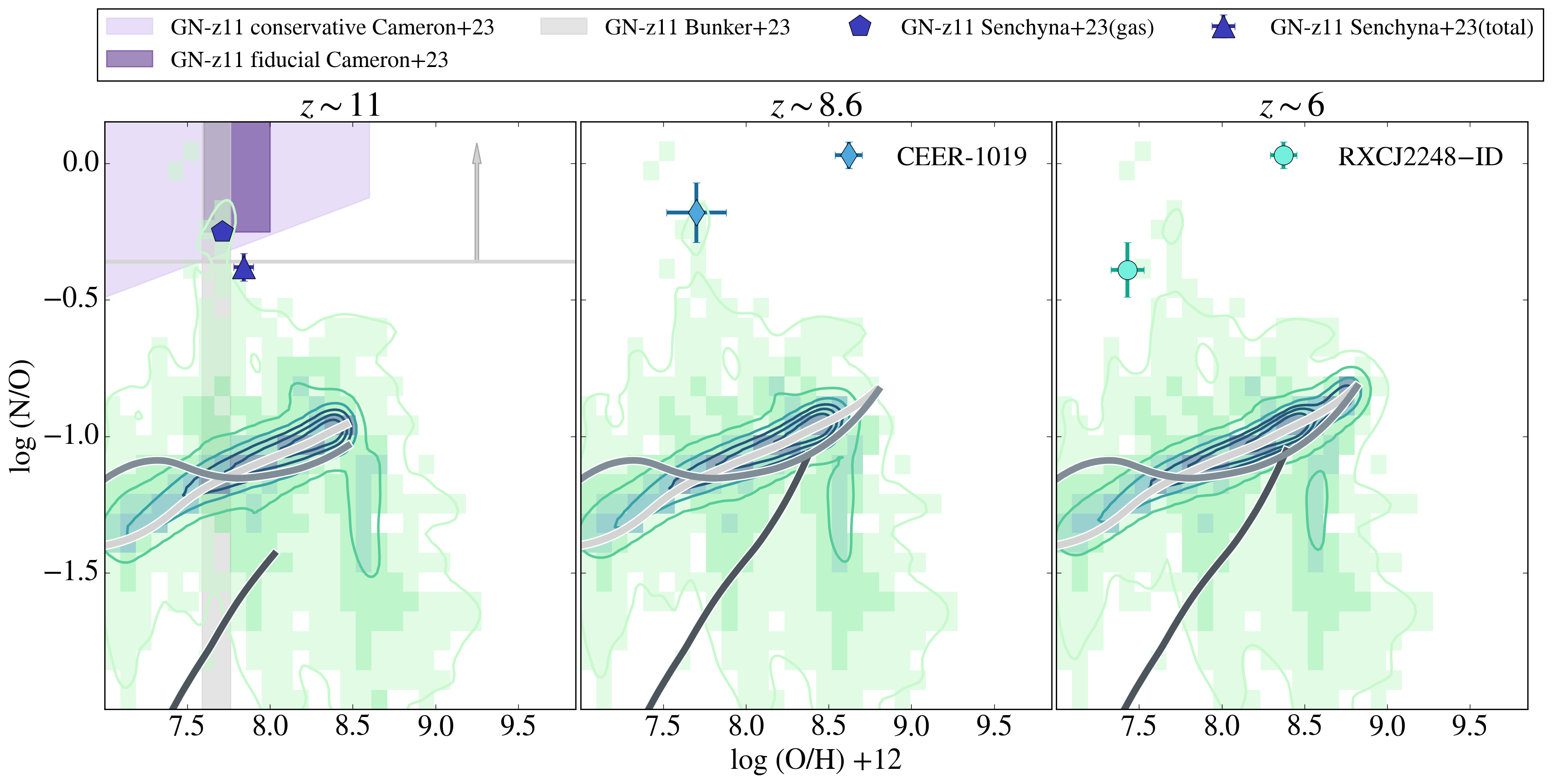}
    \caption{ 
    Theoretical log(N/O) vs log(O/H)+12 for a proto-MW galaxy seen at different redshifts. Density maps and contours refer to the predictions of the stochastic GCE model, while the solid lines refer to the predictions of the homogeneous GCE model. Abundances measured in GN-z11 \citep[$z =11$,][]{Bunker23,Cameron23,Senchyna2024}, CEERS-1019 \citep[$z = 8.6$,][]{MarquesChaves2024}, and RXCJ2248-ID \citep[$z=6$,][]{toppin24} are also shown for comparison (symbols and shaded areas).}
    \label{fig:6}
\end{figure*}

Recent JWST spectroscopy of high-redshift,  {metal-poor} galaxies, particularly the most luminous ones, such as GN-z11 \citep{Oesch2016} and CEERS-1019 \citep{Finkelstein17}, has revealed gas substantially enriched in nitrogen \citep[see][]{Bunker23,Cameron23,Maiolino23,MarquesChaves2024,Senchyna2024}, sparking renewed interest in the evolution of the CNO elements and prompting new inquiries into their origins.  {While N-emitters tend to have normal C content for their metallicities \citep[see][their figure~2]{Schaerer2024}, extreme C enhancement is found in a few systems \citep{deugenio2024}, which deserves special attention.} In this section, we aim to provide testable predictions for a MW-like galaxy observed at different redshifts, along with a possible interpretation of  {the high N/O (C/O) ratio measured for GN-z11 (GS-z12),} within the framework of our GCE model, including pre-enrichment by Pop~III SNe with yields dependent on the mass, internal mixing level, and explosion energy of the progenitor star \citep{Heger+woosley10}.

\subsection{MW counterparts at high redshift}
\label{highmw}

In Fig.~\ref{fig:6}, we display density maps in the plane $\rm log(N/O)$--$\rm  log(O/H) + 12$ for a MW-like galaxy\footnote{By MW-like galaxy, we mean a galaxy that will end its evolution with chemical and physical properties similar to our present-day MW.} seen at various redshifts. 
After the initial chemical enrichment by Pop~III SNe, which sets a wide range of initial abundances in log(O/H) and log(N/O), there is a phase of rapid chemical enrichment. The density contours show that most of the clumps that are in the process of merging to form the halo, reach values of $\rm log(N/O) \sim -1$ and $\rm log(O/H) + 12 \sim 8.5$ after only $\sim 400$~Myr $(z \sim 11)$. Observations of a MW-like galaxy at high redshift would not resolve the different knots and would thus measure those average values, which also agree with the predictions of homogeneous GCE models including massive fast rotators. The subsequent evolution proceeds slowly. Over the approximately 500 million years elapsed between $z \sim 11$ and $z \sim 6$, the average metallicity gradually increases. At $z \sim 6 $, the gas reaches $\rm log(O/H) \approx 9$ and $\rm log(N/O) \approx -0.7$, reflecting the ongoing enrichment processes and contributions from subsequent stellar generations. In Appendix~\ref{prediction MW other abundances}, we display our predictions for a proto-MW up to $z \sim 2$ and provide the maps also for the C/O and C/N abundance ratios.

Figure~\ref{fig:6} also shows the $\rm log(N/O)$ values measured in GN-z11 at $z= 10.6$ \citep{Bunker23, Cameron23, Maiolino23, Senchyna2024}, in CEERS-1019 at $z= 8.6$ \citep{MarquesChaves2024} and in RXCJ2248-ID at $z \sim 6$ \citep{toppin24}.
Note that while our predictions for the proto-MW successfully cover the observed metallicity range of $7 < \rm log(O/H) + 12 < 9$, they struggle to reproduce the high observed values of $\rm log(N/O)> -0.5$. 
This discrepancy is not unexpected: these systems are significantly brighter and more massive than what is expected for the MW progenitor at the same redshifts. For instance, at $z \approx 11$, the surface stellar density predicted for a proto-MW is $\Sigma_{\star} \sim 430$~M$_{\odot}$ pc$^{-2}$, whereas for GN-z11, it is approximately four times higher, $\Sigma_{\star} \sim 1650 $~M$_{\odot}$ pc$^{-2}$ \citep{Tacchella23}. On the other hand, our results indicate that it is possible to achieve values as high as $\rm log(N/O) > -0.5$ in some regions of a forming MW analog within the metallicity range of $ \rm  7.5 < log(O/H) + 12 < 8.0$. 
 As already discussed in Sect.~\ref{different energies}, such high N/O values are due to the contribution of faint Pop~III SNe (see Fig.~\ref{fig:no_ prob}). Accounting for the effects of stellar rotation in $Z = 0$ massive star models would likely lead to even higher N/O ratios \citep{Meynet06,Limongi+18}, but these are not implemented in the yields of \cite{Heger+woosley10} for Pop~III stars adopted in this work.

\subsection{Extreme abundance ratios at high redshift}
\label{gnz11}

 {In this section, we deal with compact metal-poor, high-redshift systems that display C/O or N/O abundance ratios at variance with what is expected for their metallicities.}

\subsubsection{GN-z11}
 
\begin{figure*}
\centering
\includegraphics[width=\textwidth]{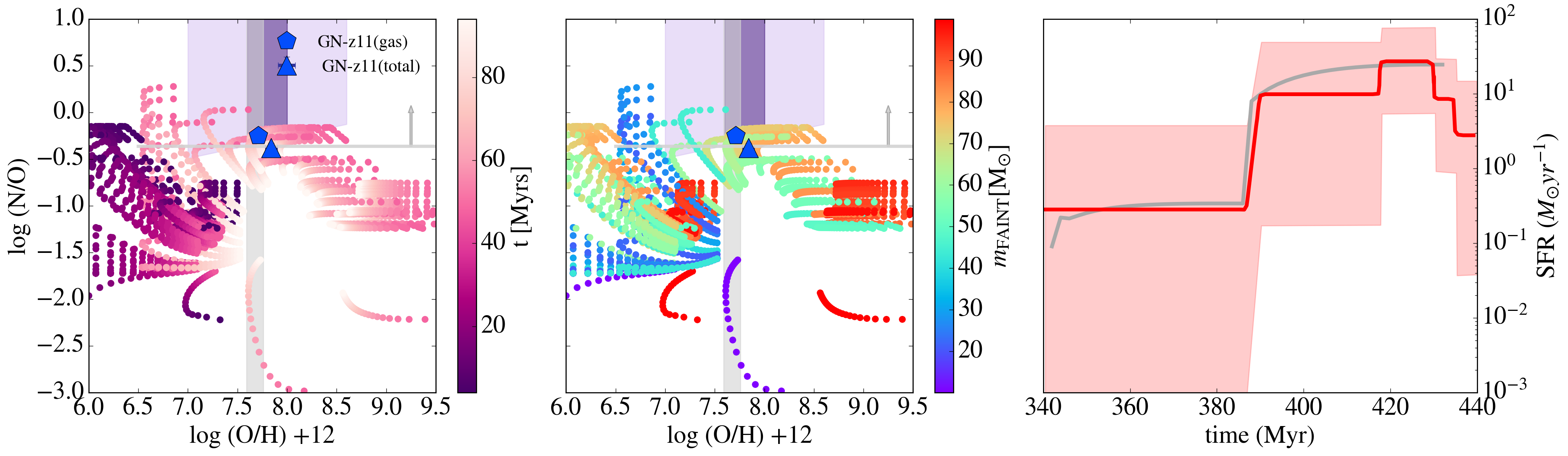}
    \caption{ {Left and middle panels: predicted log(N/O) vs log(O/H)+12 for GN-z11. Different evolutionary tracks represent different Pop~III star-forming clumps, which merge into the main-branch of the galaxy. The tracks are colour-coded in time elapsed since the beginning of star formation (left) and stellar mass (middle). The light-purple trapezoid, the dark-purple rectangle and the grey stripe represent the conservative and fiducial results for the abundances of GN-z11 by \cite{Cameron23}. The blue points are the photoionisation modelling results for GN-z11 by \cite{Senchyna2024}. Right panel: star formation history of GN-z11 according to \citet[][red line and shaded area]{Tacchella23} and as adopted in the model (grey line).}
    }
    \label{fig:9}
\end{figure*}

Figure~\ref{fig:9} presents the elemental abundance ratio $\rm log(N/O)$  {as a function of $\rm log(O/H)+12$ as a proxy for metallicity in the ISM of GN-z11, predicted by a model tailored to this galaxy (see Sect.~\ref{high redshift galaxy}).} The different traces illustrate the chemical evolution of different clumps, each harbouring one Pop~III polluter at the beginning of the simulation, at fixed density so as to reproduce the observed properties of GN-z11. In the left panel, the traces are colour-coded based on the time elapsed from the beginning  {of star formation in the model, whereas in the middle panel} different colours indicate different masses of the Pop~III stars that pre-enriched the ISM.  Note that we specifically selected models where the ISM was pre-enriched by faint Pop~III SNe, as Pop~III SNe with higher explosion energies are not able to achieve log(N/O) values that align with the observational data (see Fig.~\ref{fig:no_ prob}).  {The stellar mass and star formation rate (Fig.~\ref{fig:9}, right-hand panel) predicted by the model at the time of the observations, 440~Myr after the Big Bang, are $8.8 \times 10^8$~M$_\odot$ and 25 M$_\odot$~yr$^{-1}$, respectively, which perfectly align with the data (by construction).}
 
 {Two distinct sets of evolutionary tracks are visible, each corresponding to different star formation bursts. The dark purple tracks in Figure \ref{fig:9} (left panel) represent the chemical evolution during the initial low-level star formation activity, which lasts for approximately 40 Myr from the onset of galaxy formation. The lighter-coloured tracks correspond to the second star formation burst, which begins 40 Myr after the beginning of the simulation and evolves for approximately 60 Myr. Both bursts are fed by infall of gas of primordial chemical composition and initiated by pre-enrichment from Pop~III stars, which establishes the starting points of the $\rm log(N/O)$ vs $\rm log(O/H) + 12$ tracks.}

 {In the first star-formation episode, we observe three distinct chemical enrichment paths, each directly linked to the mass of the Pop~III SN that pre-enriched the ISM. In the first path, the trace starts at low values of $\rm log(N/O)\lesssim -1$ and $\rm log(O/H) + 12 < 6.0$, gradually increasing in $\rm log(N/O)$ over time. The starting point of the track reflects the enrichment from low-mass Pop~III stars ($m_{\rm Pop \: III} < 40$~M$_{\odot}$), as shown in the middle panel of Fig.~\ref{fig:9}, while the subsequent increase in log(N/O) and slight decrease in $\rm log(O/H) + 12$ is due at first to the continuous infall of pristine gas and then to the overwhelming contribution of massive Pop~II stars.
The second path involves Pop~III SNe with masses in the range 40--80~M$_{\odot}$ that pre-enrich the ISM to  $\rm \log(O/H) + 12 < 7.0$, and $\rm log(N/O) > -1.5$ values. Following this initial phase, the $\rm log(N/O)$ ratio rapidly declines due to significant oxygen production from subsequent Pop~II SNe, to slightly increase later mostly because of primary nitrogen production from Pop~II massive stars boosted by rotation.
Finally, the third path begins with pre-enrichment from massive Pop~III SNe ($m_{\rm Pop \: III}> 80$~M$_{\odot}$), resulting in high $\rm log(N/O) \sim -1$ values and $\rm 7.0 <\log(O/H) + 12 < 7.5$. This path shows an initial decreasing trend in metallicity (with an approximately constant $\rm log(N/O)$ value), followed by a subsequent increase. The initial decline in $\rm log(O/H)$ is attributed to the dilution due to the infall of pristine gas, followed by a chemical enrichment path similar to the one of the previous scenario, predominantly driven by the contribution of successive generations of Pop~II stars.}

 {As the second burst begins, fueled by infall of gas of primordial chemical composition,  a second episode of pre-enrichment by Pop~III stars occurs which, together with the cumulative enrichment from the first burst, establishes the starting point of the new evolutive track. Following this initial phase, the chemical evolution is predominantly driven by Pop~II star formation, dominating the later stages of evolution. 
This path shows an initial decreasing trend in metallicity with an approximately constant $\rm log(N/O)$ value. The initial decline in $\rm log(O/H)$ is attributed to the dilution due to the infall of pristine gas, followed by chemical enrichment predominantly driven by the contribution of successive generations of Pop~II stars.}

\begin{figure*}
\centering
\includegraphics[width=0.8\textwidth]{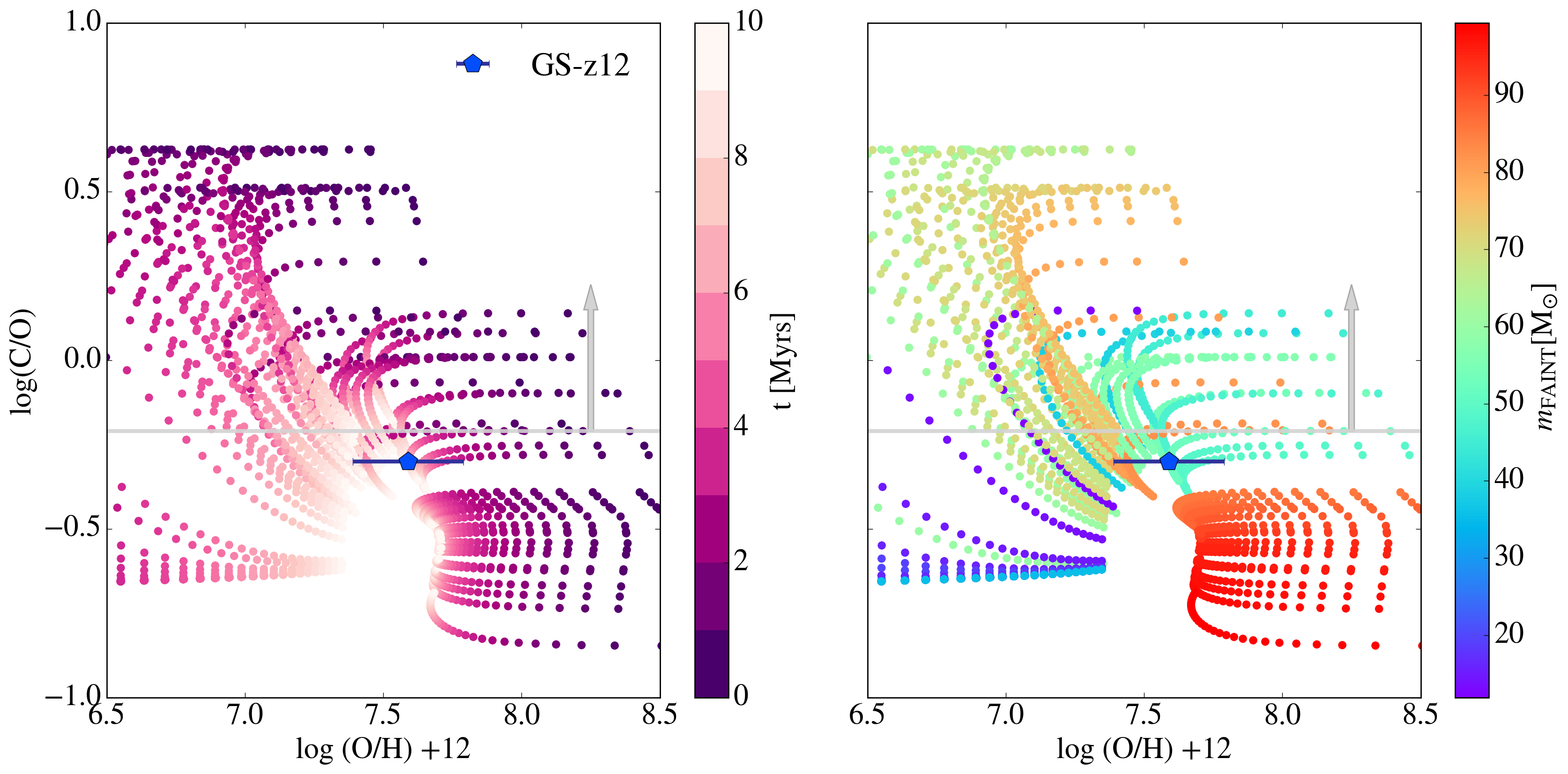}
    \caption{  {Predicted log(C/O) vs log(O/H) + 12 for GS-z12. The evolutionary tracks represent different PopIII star-forming clumps that merge into the main branch of the galaxy. The tracks are colour-coded to indicate the time elapsed since the beginning of star formation (left panel) and the PopIII stellar mass (right panel). Also shown is the measured log(C/O) abundance for GS-z12, as derived by \citet{deugenio2024}. The blue marker represents the estimate obtained through a Bayesian modeling of the spectrum, while the grey stripe marks the lower limit derived using the photoionization models of \citet{Gutkin+16}. }
    }
    \label{fig:gsz12}
\end{figure*}

 {The $\rm log(N/O)$ values predicted by our GCE model intercept the value observed in GN-z11 within the uncertainties after $100$~Myr from the beginning of galaxy formation. Setting the start of star formation at 340~Myr after the Big Bang in the model makes the predicted abundance to agree with the observed one at $z \simeq$~10.6. This suggests that the extreme N-enhancement observed at high redshift can be obtained by accounting for a pre-enrichment of the ISM from massive Pop~III stars with $m_{\rm Pop \: III} \ge$ 50--85~M$_{\odot}$ ending their lives as faint SNe. Notably, at the epoch of the observations the predicted star formation rate is $26$ M$_{\odot} \: \rm  yr^{-1}$, in agreement with the observed value, $\sim 21^{+22}_{-10}$ M$_{\odot}$ yr$^{-1} $, and the average stellar age is 20~Myr, which is comparable to the estimated value of $\sim 24^{+20}_{-10}$~Myr \citep{Tacchella23}.}

In conclusion, we hereby propose an alternative scenario to reproduce the observed high log(N/O) ratio observed in GN-z11 (and in other high-redshift extreme N-emitters) in which, before being observed, the galaxy underwent pre-enrichment by massive faint Pop~III SNe.

\subsubsection{GS-z12}

 {Figure \ref{fig:gsz12} shows the elemental abundance ratio $\rm log(C/O)$ as a function of $\rm log(O/H) + 12$ in the ISM of GS-z12, based on predictions from a model specifically designed for this galaxy. In the case of GS-z12, a single star formation burst is considered (see Sect.~\ref{high redshift galaxy}). The evolutionary tracks are colour-coded similarly to those in Fig.~\ref{fig:9}, with the colours indicating either the time elapsed since the onset of star formation (Fig.~\ref{fig:9}, left panel) or the stellar mass of the Pop~III progenitors (Fig.~\ref{fig:9}, right panel). Similar to what is found for GN-z11, distinct evolutionary tracks are clearly visible, primarily shaped by the masses of the Pop~III stars that pre-enrich the ISM.}

 {Three sets of evolutionary tracks can be recognized. In the first set, the tracks start at low values of $\rm log(C/O) \approx -0.5$ and $\rm log(O/H) + 12 < 6.5$, showing a steady decrease of $\rm log(C/O)$ as time progresses. This initial phase is driven by enrichment from low-mass Pop~III stars ($m_{\rm Pop : III} < 20$~M$_{\odot}$) that produce significant amounts of carbon with minimal oxygen contribution. As Pop~II stars enter the scene, $\rm log(C/O)$ decreases while the oxygen abundance increases, reflecting the dominant contribution of Pop~II SNe during later evolutionary phases.}

 {The second set of tracks involves a pre-enrichment by Pop~III stars with masses between 40 and 60 M$_{\odot}$. These stars enrich the ISM to high metallicities, $\rm log(O/H) + 12 \sim 8.5$, and $-0.2 \lesssim \rm log(C/O) \lesssim 0.1$. The metallicity initially decreases, while $\rm log(C/O)$ remains constant, due to the contribution of the primordial infalling gas, which dilutes the metals without affecting their abundance ratio. Subsequently, $\rm log(C/O)$ decreases due to the contribution to the enrichment by massive Pop~II SNe.}

 {The third path corresponds to Pop~III stars with masses between 60 and 80 M$_{\odot}$. These massive stars pre-enrich the ISM to low metallicities, $\rm log(O/H) + 12 \lesssim 7.5$, while $\rm log(C/O)$ reaches high values, exceeding 0.25 dex. After an initial phase in which the infall of pristine gas dilutes all the metals in the same way, the ISM becomes enriched in oxygen from Pop~II SNe and the $\rm log(C/O)$ ratio begins to decrease, while $\rm log(O/H) + 12$ steadily increases.}

 {The fourth evolutionary path is initiated by the most massive Pop~III stars ($m_{\rm Pop : III} > 80$~M$_{\odot}$), leading to initial $\rm log(C/O) \lesssim -0.4$ and high metallicity, $\rm log(O/H) \sim 8.5$. As the track progresses, there is a decrease in $\rm log(O/H) + 12$, again due to dilution from infalling pristine gas, followed by an enrichment phase dominated by Pop~II stars, which gradually increases the oxygen abundance. The $\rm log(C/O)$ values remain relatively stable, reflecting the fact that the most massive Pop~III stars exploding as faint SNe are characterizes by C/O ratios in their ejecta that resemble those of Pop~II CCSNe (see Fig.~\ref{fig:co_ prob}).}

 {Our inhomogeneous model predicts $\rm log(C/O)$ values that match the one observed in the ISM of GS-z12 after a 10-Myr long star formation burst. At the time of observation, the predicted star formation rate is approximately $1.6$~M$_{\odot} \rm yr^{-1}$, in good agreement with the observed one. The average stellar age is $\sim 10$~Myr and the stellar mass is log($M_{\star}$/M$_\odot$)~= 7.60, which also agrees with the observational estimate.}

 {In conclusion, the elevated $\rm log(C/O)$ ratio of GS-z12 can be attributed to early carbon production by Pop~III stars of masses 50--70 M$_{\odot}$, followed by later oxygen enrichment from Pop~II stars.}

\section{Discussion}
\label{discussion}

\subsection{Milky Way galaxy}

The contribution of Pop~III  stars is often neglected, or considered negligible, in chemical evolution studies of the MW. 
For instance, \cite{Ballero2006}, using the yields of \cite{Woosley95} for Pop~II SNe and those of \cite{Heger2002} for PISNe, conclude that the effects of Pop~III stars on the Galactic evolution of CNO elements are negligible, if only one or two generations of such stars are formed.

In our study, there are two significant differences compared to \cite{Ballero2006} that lead to different conclusions. First, in this work we consider the contribution of Pop~III SNe with a wide range of explosion energies, from low-energy, faint SNe to highly energetic hypernovae \citep{Heger+woosley10}, which results in theoretical [X/Fe] abundance ratios covering a broader spectrum. Second, we include a stochastic star-formation component that is fundamental to adequately describe the initial phases of galaxy formation \citep[see][]{Rossi+21}. This approach avoids using IMF-averaged yields and instead considers the contributions of individual Pop~III SNe during the early stages of the MW evolution. Averaging the yields over the IMF tends to smooth out the unique chemical signatures of individual Pop~III SNe, resulting in predicted [X/Fe] ratios that are systematically lower (or higher) than the most extreme values.

\noindent By incorporating these factors, our approach reevaluates the significant role that Pop~III SNe could play when their varied explosion energies and stochastic formation processes are taken into account.

Our findings are consistent with those of other recent studies. For example, \cite{Vanni23}, utilizing a general parametric model for early metal enrichment, illustrate how the star-to-star scatter in abundance ratios among low-metallicity halo stars is influenced by the different initial masses and SN explosion energies of Pop~III SNe. Conversely, at higher metallicities, the dispersion in abundance ratios decreases with the increased contribution from typical Pop~II stars. Similarly, \cite{ioanna2023}, employing a cosmological chemical evolution model for the MW, emphasize the necessity of incorporating faint  Pop~III SNe to accurately reproduce the carbon-enriched stars at $\rm [Fe/H] < -2$ \cite[see also][]{Rossi23}.

The scenario proposed in this work is not intended to be exclusive. We do not exclude the possibility that other sources with different properties may have contributed to the chemical enrichment of the Galaxy at low metallicities.  For instance, \cite{Chiappini05} investigated the impact of stars with metallicities $Z = 10^{-8}$ and a wide range of initial rotational velocities \citep{Hirschi04} on the early MW chemical evolution. Their study concluded that scenarios involving fast rotators at low metallicities are necessary to explain the observed high N abundances in halo stars.
Our findings aim to complement such existing research.

\subsection{C- and N-enhanced high-redshift systems}
Different scenarios have been proposed to explain the extreme N/O abundance ratios measured in high-redshift systems, such as GN-z11 and CEERS-1019 \citep{Cameron23,MarquesChaves2024}. 
\cite{Kobayashi&Ferrara24} proposed a dual-burst model emphasizing the crucial role of Wolf-Rayet (WR) stars as N producers during the secondary burst phase. \cite{Prantzos17} and \cite{Gratton19} linked chemical peculiarities in globular clusters to the extreme patterns observed in galaxies like GN-z11.  {On the same line, \citet{dantona2023} proposed that the high N content of GN-z11 could be due to the formation of second-generation stars from pristine gas and asymptotic giant branch stellar ejecta in a massive proto-globular cluster.} \cite{Charbonnel23} and \cite{Nagele23} explored the impact of metal-enriched stars, especially those within the mass range $10^3$--$10^5$~M$_{\odot}$, on early galaxy chemistry. \cite{Isobe23} investigated chemical abundances in 70 galaxies, correlating them with WR stars, supermassive stars (SMS), and tidal disruption events. 

At the same time, various studies have analysed different efficient N production channels in stars. For instance, \cite{Meynet06} and \cite{Choplin+18} delved into the effects of rapid rotation on massive stars, while \cite{Heger2002} and \cite{Takahashi18} explored PISNe. \cite{Nandal24Massive} investigate the impact of primordial SMS with masses in the range 500--9000~M$_{\sun}$ and 
\cite{Vink23} highlights the potential of very massive stars (100--1000~M$_{\sun}$) as efficient N polluters. \cite{MarquesChaves2024} consider the contribution of both WR stars and SMS with masses over 1000 M$_{\sun}$.
Finally, \cite{Nandal24PopIII} emphasize the unique contribution of Pop~III and extremely metal-poor stars in enhancing nitrogen abundance in high-redshift galaxies, considering different rotation velocities for stars in the mass range 9--120~M$_{\sun}$.

While a comprehensive interpretation remains elusive, our study contributes to the intricacies of this complex scenario, providing an alternative perspective on high-redshift N-emitters. Utilizing our stochastic GCE model benchmarked against MW data first and, then, tailored to GN-z11, we demonstrate that it is possible to achieve $\rm log(N/O) > -0.5$ through enrichment mechanisms involving low-energy ($\rm E_{SN} \le 0.6 \times 10^{51}$), faint SNe with masses exceeding 50~M$_{\odot}$.
While these pristine stars are expected to be rare, their contribution to the chemical enrichment within metal-poor environments could become relevant, should the galaxy-wide stellar IMF (gwIMF) be shifted towards higher masses \citep{Rossi+21, Ioanna24}. Indeed, evidence for a top-heavy IMF in starbursts has accumulated over the years \citep[e.g.,][]{Jerabkova2018,Schneider2018,Zhang2018,Zinnkann2024}, making our findings particularly intriguing.

 {As for the class of the rarer extreme C-emitters \citep[see][their figure~2]{Schaerer2024}, we study the emblematic case of GS-z12. Also in this case, massive Pop~III stars that explode as faint SNe are found to play a key role in reproducing the observed high C/O ratio at low metallicity, notwithstanding the lowest stellar mass reached by GS-z12.} 

\section{Conclusions}
\label{conc}
To explore the earliest phases of chemical enrichment in galaxies, we 
implement a stochastic star-formation component and detailed Pop~III nucleosynthesis prescriptions in a GCE model that already reproduces satisfactorily well the main observed properties of our Galaxy (see Sects.~\ref{MW model} and \ref{model implementations}). By comparing our model predictions with a homogeneous stellar sample of unmixed stars (Sect.\ref{sampleale22}), whose chemical abundances reflect the composition of their natal birth clouds, we first demonstrate that the scatter in CNO abundances and abundance ratios observed in the MW halo can be largely attributed to the contributions of Pop~III SNe with different masses and explosion energies. Then, we extend the model to the high-redshift universe and predict the chemical abundance ratios for a proto-MW galaxy seen at different redshifts (Sect.~\ref{highmw}). Finally, we adapt the model and provide an alternative interpretation of the chemical abundances observed in GN-z11  {and GS-z12} (Sect.~\ref{gnz11}). Our key findings can be summarized as follows:

\begin{itemize}
    \item[(i)] The spread in chemical abundances (C, N, O, and Mg) of our stellar sample  
    is not related to the sum of distinct, tighter patterns due to the accretion of different substructures in individual mergers. At the low metallicities covered by our sample, the observed scatter is likely due to intrinsic processes.
    \item[(ii)] To reproduce the observed spread in C and N, it is necessary to consider stochastic chemical enrichment driven by Pop~III stars of different masses and explosion energies, ranging from low-energy faint SNe to hypernovae. Specifically, low-energy Pop~III SNe (faint and CCSNe) impact on the dispersion at $\rm [C/Fe] > 1$ and $ \rm[N/Fe] > 0$, 
    whereas energetic Pop~III SNe contribute at  $\rm [C/Fe] < 0.5$ and $ \rm [N/Fe] < 0$, and higher [Fe/H] values.
    \item[(iii)]  High $\rm log(N/O) > -1 $ are predicted to be observed in MW halo stars, primarily driven by chemical enrichment from faint Pop~III SNe, which are the only kind of Pop~III SNe  \citep[at least, when adopting the yields of][]{Heger+woosley10} capable of achieving such high N enrichment levels.
    \item[(iv)] The chemical enrichment of MW-like galaxies is predicted to be very rapid at high redshift ($z \sim 11$) and followed by a more gradual enrichment at later stages ($z \sim 2$, see the Appendix~\ref{prediction MW other abundances}). We anticipate a large scatter in CNO abundance ratios: $-2 < \rm log(N/O) < -0.2$, $-1.5 < \rm log(C/O) < 0.2$, and $ 0 < \rm log(C/N) < 1.7 $ (see Fig.~\ref{fig:7}).
    \item[(v)] We propose an alternative interpretation of the extremely high N abundance measured in the low-metallicity, compact system GN-z11, namely, we propose that, before being observed, the galaxy was chemically enriched by massive, faint Pop~III SNe, resulting in $\rm log(N/O) > -0.25$.
    \item[(vi)]  {Pre-enrichment by faint Pop~III SNe could also be the key to explain the high C/O ratios observed in young low-metallicity systems, such as GS-z12.}
\end{itemize}

In conclusion, this study has highlighted the importance of the contribution of Pop~III stars to the early phases of galactic evolution. 
Our results emphasize the complexity of chemical enrichment processes in the early stages of galaxy formation and underscore the need for further efforts to better understand the contribution of Pop~III stars. Future work should integrate more detailed observational data and more sophisticated theoretical models to reduce the remaining uncertainties and provide a clearer picture of chemical enrichment mechanisms in the primordial universe. In this respect, the full exploitation of the capabilities of space telescopes such as JWST and Euclid will be crucial.
\section{Data availability}
Tables \ref{tab spectro} and \ref{tab orbits} are only available in electronic form at the CDS via anonymous ftp to cdsarc.u-strasbg.fr (130.79.128.5) or via http://cdsweb.u-strasbg.fr/cgi-bin/qcat?J/A+A/.
\begin{acknowledgements}
We thank Piercarlo Bonifacio and Elisabetta Caffau for providing useful suggestions and commenting on an earlier version of this paper. 
This research is part of the project \emph{"An in-depth theoretical study of CNO element evolution in galaxies"} that is supported by the Italian National Institute for Astrophysics (INAF) through Finanziamento della Ricerca Fondamentale, Theory Grant Fu.~Ob.~1.05.12.06.08 (PI: D.~Romano). DR and AM acknowledge financial support from PRIN-MIUR-22, project \emph{"LEGO -- Reconstructing the building blocks of the Galaxy by chemical tagging"} (P.I. A. Mucciarelli). Based on observations collected at the ESO-VLT under programmes 68.D-0546, 69.D-0065, 70.D-0009, 71.B-0529, 072.B0585, 074.B-0639, 076.D-0451, 078.B-0238, 090.B-0605, 092.D-0742, 099.D-0287, 0103.D-0310, 0104.B-0487, 0104.D-0059, 165.N-0276, 169.D-0473, 170.D-0010, 281.D-5015, and 380.D-0040, at the La Silla Observatory under the programme 60.A-9700, at the Magellan telescope under programmes CN2017A-33 and CN2017B-54, and on data available in the ELODIE archive. This work has made use of data from the European Space Agency (ESA) mission Gaia (https://www.cosmos.esa.int/gaia), processed by the Gaia Data Processing and Analysis Consortium (DPAC, https://www.cosmos.esa.int/web/gaia/dpac/consortium). Funding for the DPAC has been provided by national institutions, in particular the institutions participating in the Gaia Multilateral Agreement. This work made use of SDSS-IV data. Funding for the Sloan Digital Sky Survey IV has been provided by the Alfred P. Sloan Foundation, the U.S. Department of Energy Office of Science, and the Participating Institutions. SDSS-IV acknowledges support and resources from the Center for High Performance Computing  at the University of Utah. The SDSS website is \url{www.sdss4.org}. SDSS-IV is managed by the Astrophysical Research Consortium for the Participating Institutions of the SDSS Collaboration including the Brazilian Participation Group, the Carnegie Institution for Science, Carnegie Mellon University, Center for Astrophysics | Harvard \& Smithsonian, the Chilean Participation Group, the French Participation Group, Instituto de Astrof\'isica de Canarias, The Johns Hopkins University, Kavli Institute for the Physics and Mathematics of the Universe (IPMU) / University of Tokyo, the Korean Participation Group, Lawrence Berkeley National Laboratory, Leibniz Institut f\"ur Astrophysik Potsdam (AIP),  Max-Planck-Institut f\"ur Astronomie (MPIA Heidelberg), Max-Planck-Institut f\"ur Astrophysik (MPA Garching), Max-Planck-Institut f\"ur Extraterrestrische Physik (MPE), National Astronomical Observatories of China, New Mexico State University, New York University, University of Notre Dame, Observat\'ario Nacional / MCTI, The Ohio State University, Pennsylvania State University, Shanghai Astronomical Observatory, United Kingdom Participation Group, Universidad Nacional Aut\'onoma de M\'exico, University of Arizona, University of Colorado Boulder, University of Oxford, University of Portsmouth, University of Utah, University of Virginia, University of Washington, University of Wisconsin, Vanderbilt University, and Yale University.
\end{acknowledgements}

\bibliographystyle{aa}
\bibliography{aanda}

\begin{appendix}

\section{Predictions for MW analogs at different redshifts}
\label{prediction MW other abundances}
\begin{figure*}
\centering
\includegraphics[width=0.7\textwidth]{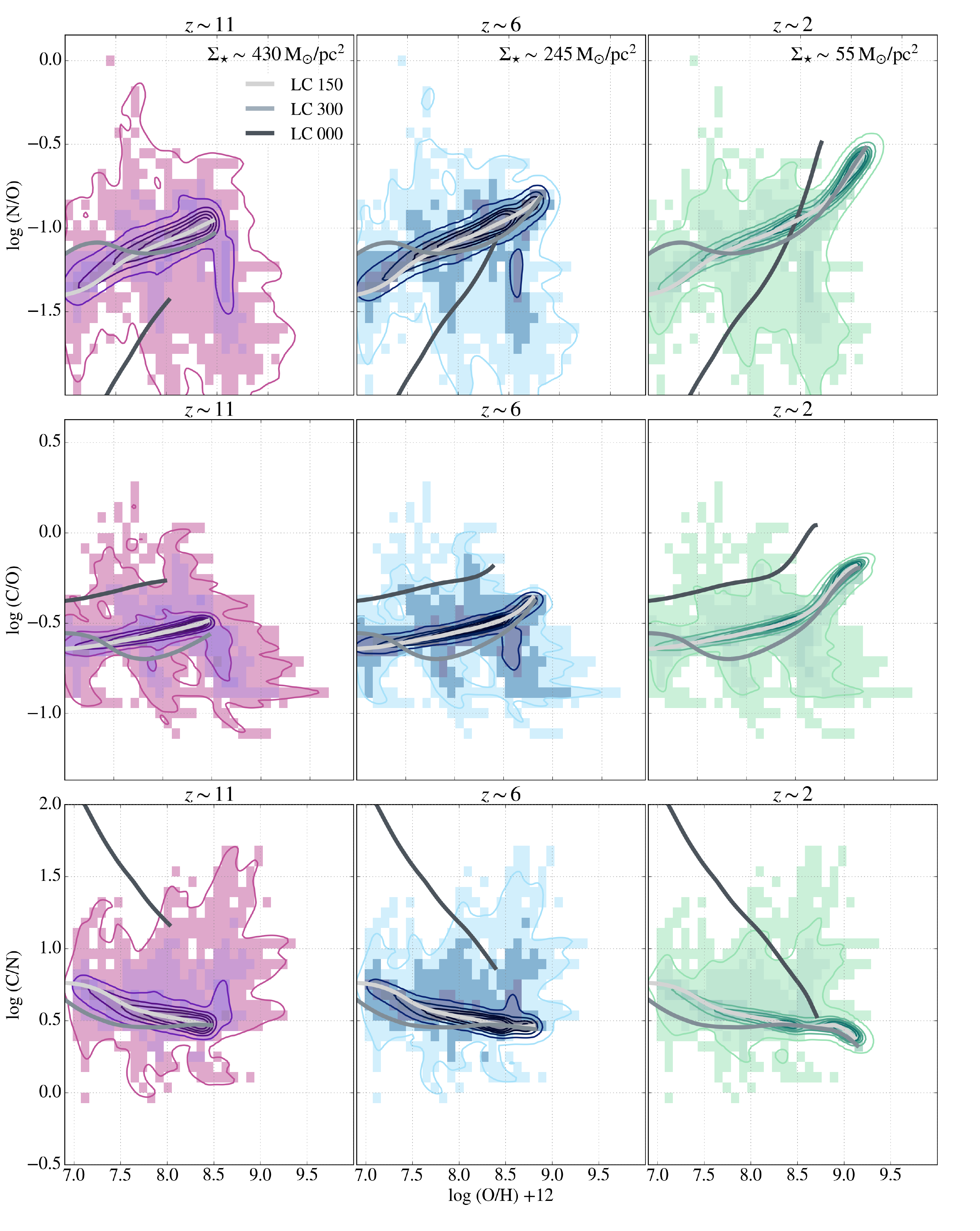}
    \caption{Chemical density maps for MW analogs seen at different redshifts. The predictions of the classic, homogeneous GCE model (solid lines, see text) are also shown.}
   \label{fig:7}
\end{figure*}

In Fig.~\ref{fig:7} we display the density maps in different planes, $\rm log(N/O)$, $\rm log(C/O)$, and $\rm log(C/N)$ versus $\rm log(O/H) + 12$, for a Milky Way-like galaxy at different redshifts, up to $z\sim 2$.
Following the initial chemical enrichment by Pop~III SNe, which establishes a wide range of initial abundances, a phase of rapid chemical enrichment occurs. The density contours indicate that most clumps merging to form the halo achieve values of approximately $\rm log(N/O) \sim -1$, $\rm log(C/O) \sim -0.5$, $\rm log(C/N) \sim 0.5$  at $\rm log(O/H) + 12 \sim 8.5$ after only about 400~Myr (at $z \sim 11$). The subsequent evolution proceeds slowly. At $z\sim 2$ the gas reaches $\rm \log(N/O) \sim -0.5 $, $ \rm \log(C/O) \sim -0.1 $, $ \rm \log(C/N) \sim 0.5 $ and $ \rm \log(O/H) +12 \sim 9.1$ reflecting the contribution of subsequent stellar generations.

%

\end{appendix}

\end{document}